\documentclass[acmtog,authorversion]{acmart}

\usepackage{booktabs} 

\usepackage[ruled]{algorithm2e} 

\SetAlFnt{\small}
\SetAlCapFnt{\small}
\SetAlCapNameFnt{\small}
\SetAlCapHSkip{0pt}
\IncMargin{-\parindent}

\acmJournal{TOG}
\acmVolume{9}
\acmNumber{4}
\acmArticle{39}
\acmYear{2010}
\acmMonth{3}

\setcopyright{usgovmixed}

\acmDOI{0000001.0000001_2}

\setcitestyle{square}
\usepackage{multirow}
\newcommand*\mean[1]{\bar{#1}}

\newcommand{\myColor}[1]{\textcolor{black}{#1}}

\begin{document}
\title{Lightweight Structure Design Under Force Location Uncertainty} 
\author{Erva Ulu}
	\affiliation{%
  	\institution{Carnegie Mellon University}}
  	\email{eulu@cmu.edu}
\author{James McCann}
	\affiliation{%
	\institution{Carnegie Mellon University}}
	\email{	jmccann@cs.cmu.edu}
\author{Levent Burak Kara}
	\affiliation{%
	\institution{Carnegie Mellon University}}
	\email{lkara@cmu.edu}

\renewcommand\shortauthors{Ulu, E. et al}

\begin{abstract}
We introduce a lightweight structure optimization approach for problems in which there is uncertainty in the force locations. Such uncertainty may arise due to force contact locations that change during use or are simply unknown a priori. Given an input 3D model, regions on its boundary where arbitrary normal forces may make contact, and a total force-magnitude budget, our algorithm generates a minimum weight 3D structure that withstands any force configuration capped by the budget. Our approach works by repeatedly finding the most critical force configuration and altering the internal structure accordingly. A key issue, however, is that the critical force configuration changes as the structure evolves, resulting in a significant computational challenge. To address this, we propose an efficient critical instant analysis approach. Combined with a reduced order formulation, our method provides a practical solution to the structural optimization problem. We demonstrate our method on a variety of models and validate it with mechanical tests. 
\end{abstract}

%
%


\setcopyright{acmcopyright}
\acmJournal{TOG}
\acmYear{2017}\acmVolume{36}\acmNumber{4}\acmArticle{158}\acmMonth{7} \acmDOI{http://dx.doi.org/10.1145/3072959.3073626}

%
%




\begin{teaserfigure}
	\includegraphics[width = \textwidth]{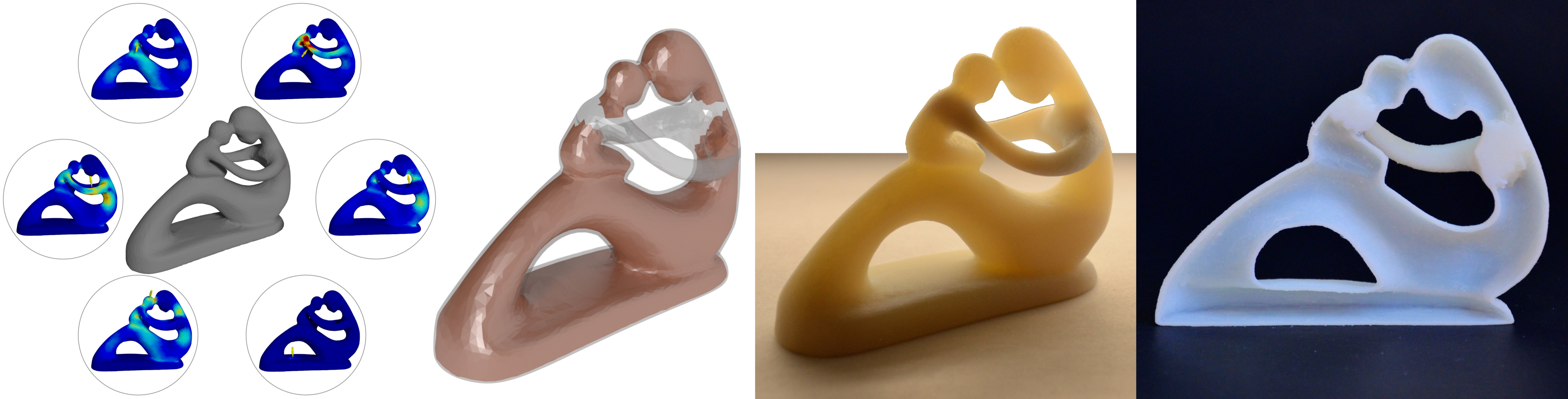}
   	\caption{We present a method for lightweight structure design for scenarios where external forces may contact an object at a multitude of locations that are unknown a priori. For a given surface mesh (grey), we design the interior material distribution such that the final object can withstand  all external force combinations capped by a budget. The red volume represents the carved out material, while the remaining solid is shown in clear.  Notice the dark material concentration on the fragile regions of the optimum result in the backlit image. The cut-out shows the corresponding  interior structure of the 3D printed optimum.}
   	\label{fig:teaser}
\end{teaserfigure}

\maketitle

\section{Introduction}

With the emergence of additive fabrication technologies, structural optimization and lightweighting methods have become increasingly ubiquitous in shape design~\cite{stava2012stress,wang2013cost,lu2014build,christiansen2015combined}. In many such methods, a common approach is to model the external forces as known and fixed quantities. In many real world applications, however, the external forces' contact locations and magnitudes may exhibit significant variations during the use of the object. In such cases, existing techniques are either not directly applicable, or require the designer to make overly conservative simplifications  to account for the uncertainty in the force configurations \cite{choi2002structural}.

We propose a new method for designing minimum weight objects when there exists uncertainty in the external force locations. Such uncertainties may arise in various contexts such as (\textit{i})~multiple force configuration problems where the object experiences a large set of known force configurations such as those arising in machinery, (\textit{ii})~unknown force configuration problems where the location of the contact points may change nondeterministically such as consumer products that are handled in a multitude of ways, or  (\textit{iii})~moving contact problems where a contact force travels on the boundary of an object; such as automated fiber placement manufacturing or cam-follower mechanisms. 

Our approach takes as input (1) a 3D shape represented by its boundary surface mesh, (2) a user-specified \textit{contact region}; a subset of the boundary where external forces may make contact, and (3) a \textit{force-budget}; a maximum cap on the total summed magnitude of the external forces at any given time instance, and produces a minimum weight 3D structure that withstands any force configuration capped by the budget (Figure~\ref{fig:teaser}).

For structural optimization with force location uncertainties, a seemingly reasonable approach would be to compute an optimal structure for every possible force configuration and select the \textit{best} structure at the end. However, this strategy fails to guarantee that the final structure (or any other optimum structure computed along the way) is safe under any force configuration other than the one it was computed for \cite{banichuk2013introduction}. Therefore, at a minimum, finding the best structure requires validating each optimum structure against all possible force configurations. Unfortunately, even this strategy does not guarantee that a solution exists within the set of computed optima.

Our approach overcomes these challenges using a \emph{critical instant analysis} which efficiently determines the most critical force contact location responsible for creating the highest stress within the current shape hypothesis. This capability enables each step of the shape optimization to efficiently determine the maximum possible stress  that can be generated under the force budget and accordingly design the material distribution against failure. Our approach preserves the outer shape through an ingrown boundary shell while optimization removes material from the inside. We do not permit structural alterations to the exterior of the object for strengthening. Hence, our approach is clearly not useful in cases where material failure occurs even in the fully solid version of the object.

Our main contributions are:

\begin{itemize}
\item a novel formulation for  structural optimization problems under force location uncertainty,
\item a method we call \textit{critical instant analysis} that identifies the critical load instant quickly,
\item a practical reduced order lightweighting method using the above two ideas.
\end{itemize}

\section{Related Work}
Our review focuses on  studies that highlight  fabrication oriented design, lightweight structure synthesis, and structural analysis, with an emphasis on approaches involving additive fabrication.

\paragraph{Fabrication oriented design} 
A large body of work has investigated automatic techniques for 3D shape design and additive fabrication subject to a variety of  functional requirements.
Recent examples include designing for prescribed deformation behaviors~\cite{bickel2010design,Skouras2013computational,schumacher2015microstructures,panetta2015elastic}, balancing models~\cite{prevost2013make}, spinnable objects~\cite{bacher2014spin} and broader methods that can handle multiple requirements~\cite{chen2013spec2fab,musialski2015reduced,christiansen2015combined}.
Our problem falls under the general category of weight-optimal structure design subject to external forces \cite{bendsoe1989optimal,wang2013cost,lu2014build,christiansen2015combined}. However, our approach addresses a more general class of problems in which the precise force locations cannot be prescribed apriori, or the structure experiences forces that can contact its surface at a multitude of locations.

\paragraph{Lightweight structure synthesis}
Cellular structure \cite{medeirosESa2015adaptive},
honeycomb-like structure \cite{lu2014build},
truss element based skin-frame structure \cite{wang2013cost},
beam element based tree-like structure \cite{zhang2015medial} generation methods
and topology optimization methods \cite{christiansen2015combined,bendsoe2003topology} are among the recent lightweight internal structure synthesis techniques that consider durability as one of the primary constraints.
However, these methods assume a prescribed static force configuration for structural design.
Although driven by similar motivations, our work addresses a more general problem of structural design under force location uncertainty. On the other hand, our formulation is also complementary in that it may facilitate the extension of these previous methods to problems involving force uncertainties. 

Langlois~\textit{et. al.}~\cite{langlois2016stochastic} performs structural optimization by predicting the failure modes of objects in real world use.
Their stochastic finite element model uses contact force samples generated by rigid body simulations to predict  failure probabilities.
They perform weight minimization while limiting the failure probability below a prescribed threshold.
While their method is applicable to scenarios where loading is stochastic in nature (such as dropping and collisions),
it is not streamlined for deterministic scenarios where the set of possible force configurations are known and no failure is tolerated for any of them. However, their method is extremely well-suited to automatically generating our contact regions, thereby allowing stochastic scenarios they consider to be addressed using structural guarantees our approach enables.

Model reduction has been used for material \cite{xu2015interactive} design, with a primary emphasis on controlling deformation behavior. Our approach is similar to traditional topology optimization methods \cite{bendsoe2003topology,lee2012stress} in that we optimize the material distribution using a fixed volumetric mesh as the parameterization. However, structural optimization under force location uncertainties introduces computational challenges that make a full dimensional analysis using the original shape parameterization to be prohibitively expensive. We are thus inspired by the above reduction method for shape synthesis, and use this in our implementation in conjunction with  our new critical instant analysis.

Musialski~\textit{et al.}~\shortcite{musialski2015reduced} introduce the idea of offset surfaces for hollowing out a solid object.
This method serves as another shape parameterization for functional optimization. In our work, we use this method to form a fixed, ingrown boundary shell, and use the remaining internal volume for shape optimization. 

\paragraph{Structural analysis} 
In structural optimization, stress and deformation analysis using Finite Element Analysis (FEA) often introduce expensive computational  bottlenecks. Simple elemental structures  such as trusses \cite{smith2002creating,rosen2007design,wang2013cost} and beams \cite{zhang2015medial} have been used to alleviate this issue. For  cases where the structure cannot be represented by these simple elements, Umetani and Schmidt~\shortcite{umetani2013cross} simplify the problem into 2D cross-sections and extend the Euler-Bernoulli model into free-from 3D objects to facilitate analysis. 

Zhou~\textit{et al.}~\shortcite{zhou2013weak} extend modal analysis used in dynamic systems (such as vibrations) to static problems to identify the potential regions of a structure that may fail under arbitrary force configurations. Our critical instant analysis builds upon this approach; we use modal analysis to determine the weak regions in a similar manner.
It allows our method to determine possible failure points based purely on geometry, \textit{i.e.}~independent of the loading.
We incorporate the weak region analysis into our structural optimization to focus on only a small region in the object to monitor the stress, thereby helping the convergence. 

In bridge (traffic load) and building (wind load) design, an equivalent uniformly distributed static load can be used to perform simple approximate analysis to account for force location uncertainty \cite{choi2002structural}. However, this approach is limited to simple geometries, making it unsuitable for our purposes.

\section{Problem Formulation}

Our design problem aims to find an optimal material distribution inside the boundary surface mesh $\mathcal{S}_0$ parametrized by the discretized volumetric mesh. Similar to topology optimization \cite{bendsoe2003topology}, material design \cite{Skouras2013computational,xu2015interactive} and microstructure design \cite{schumacher2015microstructures} approaches, each element in the discretized domain is associated with a design variable $\rho_e$ representing whether  element $e$ is full ($\rho_e = 1$) or void ($\rho_e = 0$).
To overcome the computational barriers introduced by binary variables, we adopt the common approach of allowing $\rho_e \in [0,1]$ and penalize the intermediate values during  optimization
\cite{bendsoe1989optimal}.
We assume linear isotropic materials and small deformations. The elemental stiffness matrix $\boldsymbol{K}_e$ can be related to $\rho_e$ and the stiffness matrix for base material $\boldsymbol{K}_e^{solid}$ as

\begin{equation}
\boldsymbol{K}_e = \boldsymbol{K}_e^{void} +\rho_e^ \beta (\boldsymbol{K}_e^{solid} - \boldsymbol{K}_e^{void}).
\label{Eq:SIMP}
\end{equation}

Here, $\beta$ is a penalization factor and $\boldsymbol{K}_e^{void} = \epsilon \boldsymbol{K}_e^{solid}$  is the stiffness matrix assigned to the void regions to avoid singularities in FEA. We use $\epsilon = 10^{-8}$ and $\beta = 3$.
In \eqref{Eq:SIMP}, $\boldsymbol{K}_e^{solid}$ is constant for each element and is computed as 

\begin{equation}
\boldsymbol{K}_e^{solid} = V_e \boldsymbol{B}_e^T \boldsymbol{C}_e^{solid} \boldsymbol{B}_e,
\label{Eq:elementStiffness}
\end{equation}

where $V_e$ is volume of the element, $\boldsymbol{B}_e$ is the strain-displacement matrix that depends only on the element's rest shape and $\boldsymbol{C}_e^{solid}$ is the elasticity tensor constructed using the Young's modulus and Poisson's ratio of the base material.
Given a volumetric mesh $\mathcal{V}$ with $m$ elements, one can assemble $\rho_e$ into vector $\boldsymbol{\rho} \in \mathbb{R}^m$ and construct the global stiffness matrix $\boldsymbol{K}(\boldsymbol{\rho})$ in order to determine the displacements $\boldsymbol{u}$ from $\boldsymbol{K} \boldsymbol{u} = \boldsymbol{f}$, where $\boldsymbol{f}$ is the nodal force vector. Then, the stress-displacement relationship can be written as

\begin{equation}
\boldsymbol{\sigma} = \boldsymbol{C_g} \boldsymbol{B} \boldsymbol{u}, 
\label{Eq:stressFormula}
\end{equation}

where $\boldsymbol{\sigma} \in \mathbb{R}^{6m}$ captures the unique six elements of the elemental stress tensor
and $\boldsymbol{B}$ is the assembly of $\boldsymbol{B}_e$ matrices. \myColor{Block-diagonal matrix $\boldsymbol{C}_g \in \mathbb{R}^{6m \times 6m}$ is constructed with elemental elasticity tensors $\boldsymbol{C}_e(\boldsymbol{\rho})$ on the diagonal.}
For each element, $\boldsymbol{C}_e$ can be computed analogous to $\boldsymbol{K}_e$ in \eqref{Eq:SIMP}.
While applicable to different element types, we use linear tetrahedral elements making $\boldsymbol{K}(\boldsymbol{\rho}) \in \mathbb{R}^{3n \times 3n}$, $\boldsymbol{u} \in \mathbb{R}^{3n}$, $\boldsymbol{f} \in \mathbb{R}^{3n}$ and $\boldsymbol{B} \in \mathbb{R}^{6m \times 3n}$ for a volume mesh having $n$ nodes. 

The  approach formulated in~\eqref{Eq:SIMP}-\eqref{Eq:stressFormula} is useful because it preserves the same discretization through out the optimization.
Additionally, it is amenable to model reduction presented in Section~\ref{sec:ModelReduction} for more efficient iterations (at the expense of reduced degrees freedom). 

\paragraph{Force Model} 

\begin{figure}
\centering
\includegraphics[trim = 0in 0in 0in 0in, clip, width = 0.5\columnwidth]{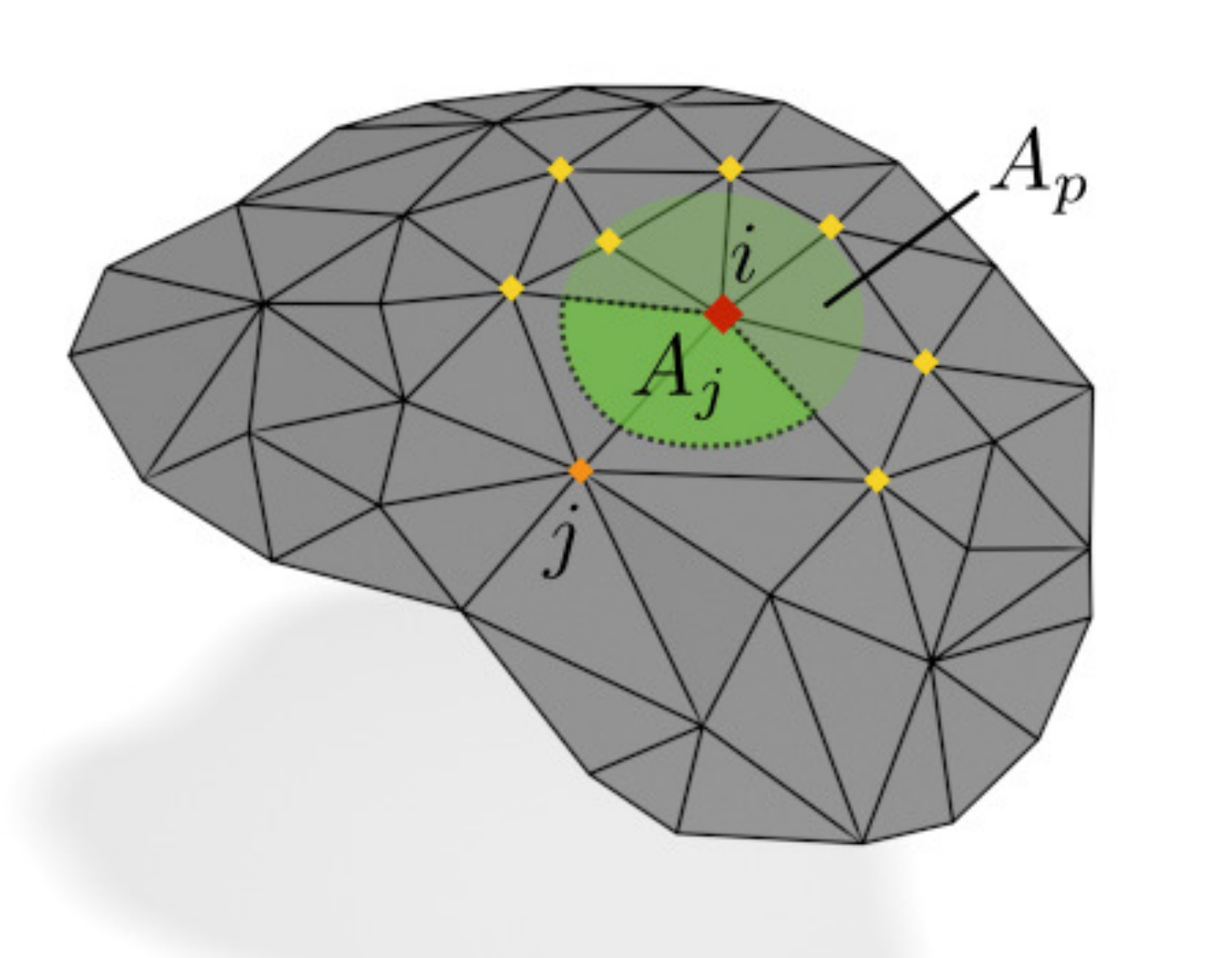}
\caption{The point force on node $i$ is spread in a circular area $A_p$. Nodal forces are computed using \eqref{Eq:areaForce}. Highlighted nodes, including $i$ and $j$, have non-zero nodal forces.}
\label{fig:areaForces}
\end{figure}

External forces are allowed to make contact within a user-specified union of contact regions $\mathcal{S}_L \subseteq \mathcal{S}_0$. To avoid stress singularities (\textit{i.e.}~unbounded stresses under a point force), we distribute the force to a small circular area $A_p$ (with radius $r_p$) around the contact point. Then, we construct the force vector $\boldsymbol{f}$ by computing the nodal forces as

\begin{equation}
\boldsymbol{p}_{j} = -P(A_j/3A_p)\boldsymbol{n}_i,
\label{Eq:areaForce}
\end{equation}    

where $\boldsymbol{p}_{j}$ is the force vector at node $j$ when a force with magnitude $P$ is applied to node $i$.
The area $A_j$ is the portion of $A_p$ covered by triangles adjacent to node $j$ and $\boldsymbol{n}_i$ is the surface normal at node $i$ (Figure \ref{fig:areaForces}).
Our approach approximates $A_p$ by intersecting the boundary mesh with a sphere of radius $r_p$ centered at node $i$. Congruent with our earlier problem description of normal contact forces only, \eqref{Eq:areaForce} assumes the force is applied  compressively along the surface normal direction. This formulation thus neglects friction and excludes forces that pull on the surface. Note, however, that most real-world contact scenarios such as handling a part or the contacts within an assembly can be modeled with compressive normal forces developing between interacting bodies.

To anchor the object in space, we require that the mesh is fixed at three or more non-collinear boundary nodes. Boundary constraints remain unchanged during  optimization. 

\begin{figure*}
\centering
\includegraphics[width = \textwidth]{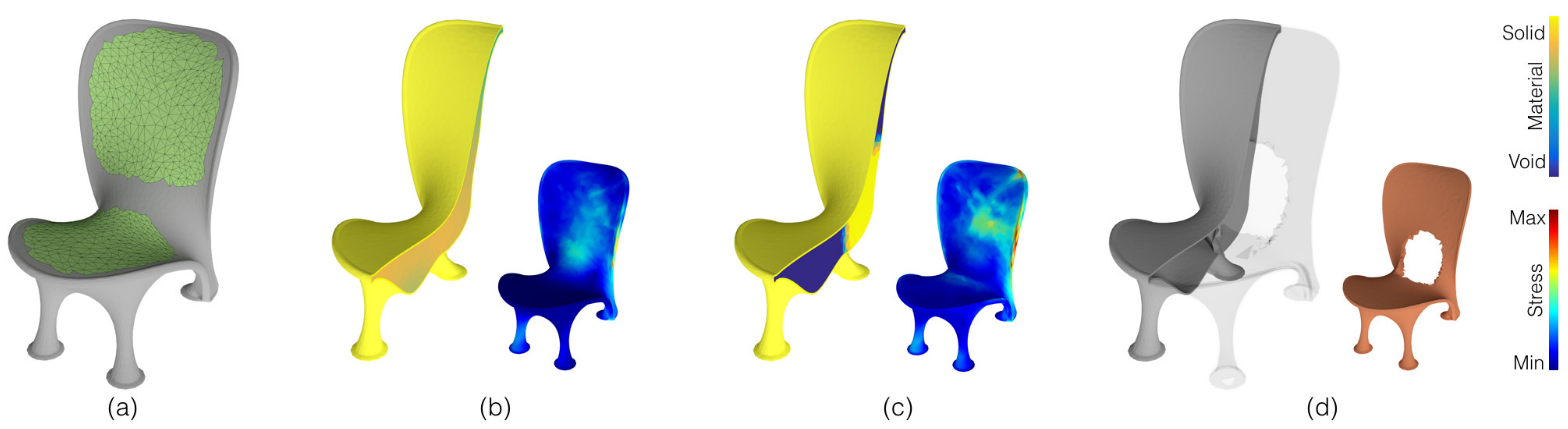}
\caption{Given a contact region (green in a), our algorithm optimizes the interior material distribution (b-c) to find the smallest weight structure (d) that can withstand all possible force configurations. In (b-c), we show the material distribution in two steps of the optimization. Inset figures illustrate the stress distributions for the most critical force instants in (b-c) and the removed material in (d).}
\label{fig:overview}
\end{figure*}

\myColor{In this work, we use the von Mises failure criterion. For linear elastic structures, the stress is a linear function of dispacement and the displacement at any given point is a linear function of the force vector. Similarly, the force vector is a linear combination of point normal forces in the contact region. Thus, the stress at a point within the structure can be determined by a superposition of each point force's contribution~\cite{hibbeler2015structural}, thereby making the von Mises stress convex in the applied force. For the force budget $f_B$, the space of allowable forces is defined by $ \| \boldsymbol{p}_{j} \| >0 $ and $ \sum \| \boldsymbol{p}_{j} \| < f_B $, \textit{i.e.}~a simplex $\mathcal{C}$ with vertices of the form $f_B$ for j'th coordinate and zero for the rest and the coordinate origin. By Rockafellar's Theorem 32.2 \cite{Rockafellar70convexanalysis}, supremum of a convex function on a convex set $\mathcal{C}$ is attained at one of the points in the set enclosed by the convex hull $\mathcal{C}$. Therefore, von Mises stress at a point will be maximized by spending the entire force budget at a specific point. The same principal holds for the maximum of the von Mises stress over the whole object as a maximum of a set of convex functions is convex. As a result, at the heart of our approach is the search for this most critical contact point given a shape hypothesis (Section~\ref{sec:Critical Instant Analysis})}


\paragraph{Optimization problem}

We tackle the following stress-constrained mass minimization problem

\begin{equation}
\begin{aligned}
& \underset{\boldsymbol{\rho}}{\text{minimize}}
& & M(\boldsymbol{\rho})  = \sum_{e=1}^{m} \rho_{e} V_e \\
& \text{subject to}
& & \boldsymbol{K}(\boldsymbol{\rho}) \boldsymbol{u}_i = \boldsymbol{f}_i \quad \forall i \in \mathcal{S}_L, \\
& & & \sigma_{cr}(\boldsymbol{\rho}) \leq \sigma_y,\\
& & & \boldsymbol{0} \leq \boldsymbol{\rho} \leq \boldsymbol{1}.
\end{aligned}
\label{Eq:optimizationProblem}
\end{equation}

Here, $\boldsymbol{f}_i$ and $\boldsymbol{u}_i$ represent the nodal force  and  displacement vectors when the external force is applied to surface node $i$.
The object fails if the maximum stress ever exceeds the yields strength $\sigma_y$.
Hence, we define the critical stress $\sigma_{cr}$ as

\begin{equation}
\sigma_{cr} = \underset{\boldsymbol{i}}{\text{max}}( \underset{\boldsymbol{e}}{\text{max}}(\enskip \sigma_e^{vm}\enskip)) \quad \forall i \in \mathcal{S}_L \enskip \text{and} \enskip \forall e \in \mathcal{V},
\label{Eq:sigmaCritical}
\end{equation}

where $\sigma_e^{vm}$ is the von Mises stress computed for element $e$. 

\section{Algorithm}

For force instant $i$, all elemental von Mises stresses $\sigma_e^{vm}$ (hence the maximum) within the object  can be computed using \eqref{Eq:stressFormula} with a single linear solve. However, finding the maximum across all possible instants require as many FEA solves as the number of instants. This number can be large, especially for  structures where $\mathcal{S}_L$ consist of many  nodes.
In such cases, computing the critical stress can be  costly, making shape optimization prohibitively expensive.
We next describe our approach to addressing this problem.

\subsection{Overview}

Figure~\ref{fig:overview} illustrates our approach. From an input 3D shape and prescribed contact regions (Figure~\ref{fig:overview}(a)),
we optimize the material distribution. At each step governed by the current material distribution, we compute the critical stress by efficiently finding the most critical force instant. We call this process \textit{critical instant analysis}. In this analysis, we reduce the search space by computing a set of \textit{force regions} (FR) over the contact region $\mathcal{S}_L$  and \textit{weak regions} (WR) within the entire structure  $\mathcal{V}$.
FRs are a subspace of the surface that are likely to contain the critical force instant. Likewise, WRs are the regions where the maximum stress is likely to occur.  We efficiently find the critical force instant within FRs using a reduced number of FEA evaluations dictated by the number of vertices within FRs. Then, optimization updates the material distribution to minimize mass (Figure~\ref{fig:overview}(b-c)).
At the end, a minimum weight structure satisfying the imposed constraints is obtained (Figure~\ref{fig:overview}(d)). Algorithm~\ref{alg:ourAlgorithm} summarizes our approach. \myColor{Note that the material distribution is updated only once at each optimization step based on the computed gradients.}


\begin{algorithm}
\myColor{
 \SetAlgoLined
 \SetKwInOut{Input}{Input}\SetKwInOut{Output}{Output}
 \Input{$\mathcal{S}_0$ and $\mathcal{S}_L$}
 \Output{Optimized structure}
 \While{Mass is reduced}{
  Compute force regions (FRs)\;  
  Compute weak regions (WRs)\;
  \For{each FR}
  {
  	Perform a hierarchical search to find largest stress at WRs\;
  }
  Choose the maximum stress across all FRs as the critical stress $\sigma_{cr}$\;
  Update material distribution $\boldsymbol{\rho}$\;
 }
 }
 \caption{Our structure optimization algorithm}
 \label{alg:ourAlgorithm}
\end{algorithm}

\subsection{Critical Instant Analysis}
\label{sec:Critical Instant Analysis}

Critical instant analysis finds the most critical force instant and the corresponding stress $\sigma_{cr}$ with an order of magnitude fewer FEA evaluations compared to a brute-force approach.

\begin{figure*}
\centering
\includegraphics[width = \textwidth]{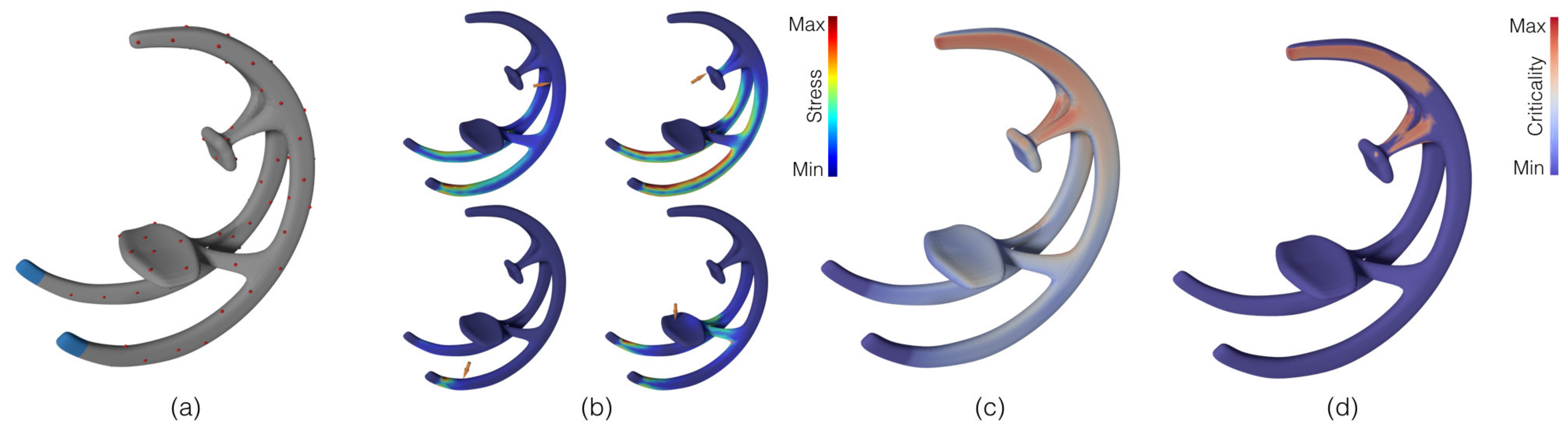}
\caption{
Given a structure represented by the material distribution, we uniformly sample a number of force instants on $\mathcal{S}_L$ (a) and perform FEAs to obtain corresponding the stress distributions (b).
We then use quadratic regression to estimate the stress distributions for the remaining force instants and construct the criticality map (c).
Areas with high criticality constitutes our force regions (d).
The blue regions in (a) represent the fixed boundary condition and the remainder of the boundary surface forms $\mathcal{S}_L$.
}
\label{fig:criticalityMapOverview}
\end{figure*}

\begin{figure}
\centering
\includegraphics[width = \columnwidth]{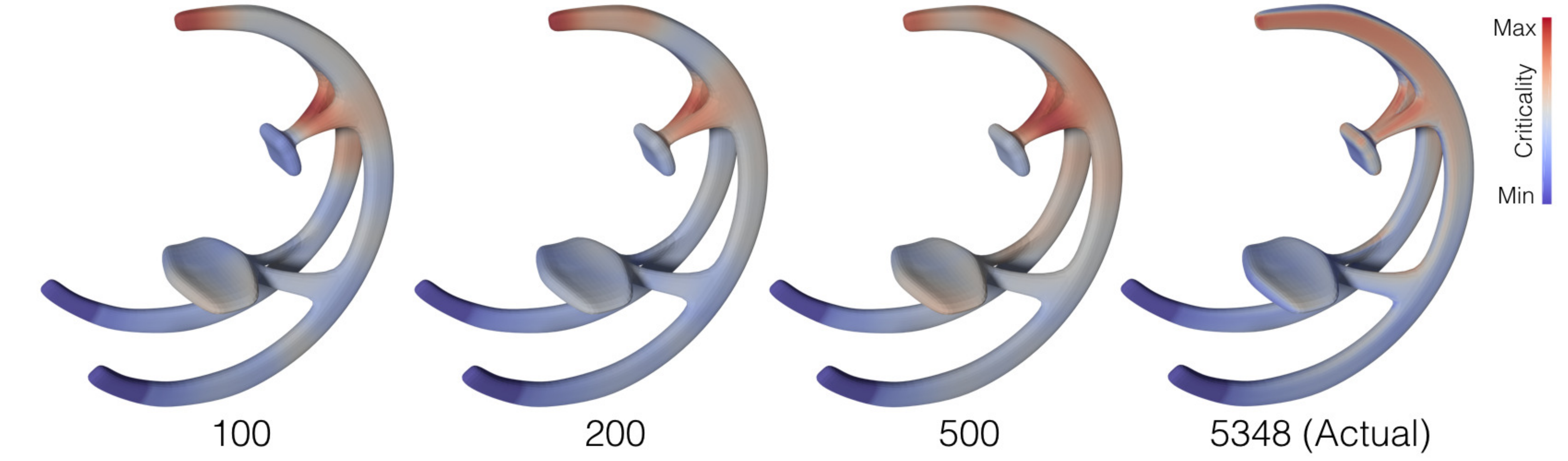}
\caption{Criticality maps as a function of the number of samples. With more samples, the estimated criticality map converges to the actual map, but at an increased cost.}
\label{fig:ActualVSEstimateCriticalityMap}
\end{figure}

\paragraph{Force Regions} 

For a structure represented by material distribution $\boldsymbol{\rho}$, certain force configurations will cause the largest stresses in the body. We compute these critical force locations as our force regions and restrict the search space for $\sigma_{cr}$ in \eqref{Eq:sigmaCritical} to a smaller space $\mathcal{S}_{fr} \subset \mathcal{S}_L$.

Figure~\ref{fig:criticalityMapOverview} illustrates our approach to compute FRs.
We start by estimating a \textit{criticality map} on $\mathcal{S}_L$, which captures the severity of the force instants. The higher stress a force instant creates, the more critical it is deemed. Thus, the criticality of a force instant is simply the maximum stress it creates in the object.

To efficiently acquire the entire criticality map, we perform FEAs  only for a small number of force instants and estimate the stress distributions for remaining force instants by learning a mapping between the nodal forces and the resulting stress distributions.
We  sample the force instants across $\mathcal{S}_L$ by performing k-means clustering using the farthest-first traversal initialization and selecting the center points of the resulting clusters as our sample instants. We use approximate geodesic distances \cite{crane2013geodesics} as the distance metric in clustering.

Suppose we have $l$ training samples and the boundary mesh $\mathcal{S}_0$ consists of $s$ nodes. In its original form, the sample force instant $\boldsymbol{f_i}$ is represented as a sparse vector of size $3n$ and the corresponding von Mises stress forms a vector of size $m$.
With a small number of training samples ($l \ll 3n$ and $l\ll m$), it is not possible to represent the relationship between two high dimensional data using a simple mapping function.
Moreover, in its sparse form, $\boldsymbol{f_i}$ is devoid of any spatial information relevant to the corresponding force instant. Hence, in this representation, two spatially proximate force instants that likely create similar stress distributions can be as distinct  as two spatially distant force instants. To  reduce the dimensionality of the force space and  establish proximity, we transform and project the sparse force vectors using the surface Laplacian. We stack the magnitudes of forces on boundary surface nodes (\textit{i.e.} $\| \boldsymbol{p} \|$ in \eqref{Eq:areaForce}) into row vectors $\boldsymbol{f'_i}$ of length $s$.
We assemble the mean centered $\boldsymbol{f'_i}$ into a $(l \times s)$ matrix $\boldsymbol{F}$ so that each row is $\boldsymbol{f'_i} - \bar{\boldsymbol{f}}$ where $\mean{\boldsymbol{f}}$ is the average of $\{\boldsymbol{f'_i}\}$.
We then compute the Laplacian basis functions $\boldsymbol{\psi_j}$ as the eigenvectors of the surface graph Laplacian $\boldsymbol{\mathcal{L}}_s \in \mathbb{R}^{s \times s}$.
We assemble the first $q$ eigenvectors to form our lower dimensional basis matrix $\boldsymbol{\Psi} = [ \boldsymbol{\psi_1}, \boldsymbol{\psi_2}, \ldots \boldsymbol{\psi_q} ]$. The lower dimensional representation of the force instants can then be written as 

\begin{equation}
\boldsymbol{F}_L = \boldsymbol{F} \boldsymbol{\Psi},
\label{Eq:forceInLaplacian}
\end{equation} 

where $\boldsymbol{F}_L$ becomes an $(l \times q)$ matrix. 

Similarly, we use principal component analysis (PCA) to project the stress data onto a lower dimensional space.
We assemble the mean centered stress vectors into an $(l \times m)$ matrix $\boldsymbol{T}$.
A PCA on $\boldsymbol{T}$ yields $(l-1)$ principal vectors of size $m$.
Then, each stress vector can be approximated by $(l-1)$ PCA weights through a linear combinations of the principal vectors 

\begin{equation}
\boldsymbol{T}_L = \boldsymbol{T} \boldsymbol{\Phi}
\label{Eq:stressInPCA}
\end{equation}

where $\boldsymbol{T}_L$ is $(l \times l-1)$ matrix storing the PCA weights for each sample in its rows and $\boldsymbol{\Phi}$ is the assembly of principal vectors.

Lower dimensionality in $\boldsymbol{F}_L$ and $\boldsymbol{T}_L$ allows us to learn a simple mapping between the two spaces with a reasonable computational cost. We have found that quadratic regression with L2 regularization performs sufficiently well for capturing the relationship between the PCA weights of the stress vectors and the reduced dimensional force vectors \myColor{such that $\boldsymbol{T}_L = \hat{\boldsymbol{F}_L} \boldsymbol{\mathcal{W}}$. Here, $\hat{\boldsymbol{F}_L}$ is $(l \times (q^2 + 3q +2)/2)$ matrix including the quadratic terms for $\boldsymbol{F}_L$. In matrix form, the coefficient matrix can be computed as}

\begin{equation}
\myColor{
\boldsymbol{\mathcal{W}} = ({\hat{\boldsymbol{F}_L}}^{T}\hat{\boldsymbol{F}_L}+r \boldsymbol{I})^{-1}({\hat{\boldsymbol{F}_L}}^{T}\boldsymbol{T}_L)
}
\end{equation}

\myColor{where $r$ is a small number controlling the importance of the regularization term.} Using this map, we estimate the criticality of a new force instant by computing the stress distribution it creates through the quadratic map, and extracting its maximum. Note that the estimated stress vectors are only an approximation of the actual values, thus cannot be used directly for $\sigma_{cr}$. However, they provide strong guidance in estimating the location of the most critical force instant. We thus use the synthesized criticality map to determine the force regions (Figure~\ref{fig:criticalityMapOverview}(d)). The connected components of high criticality areas in $\mathcal{S}_L$ comprise our force regions.

\begin{figure}
\centering
\includegraphics[width = 0.9\columnwidth]{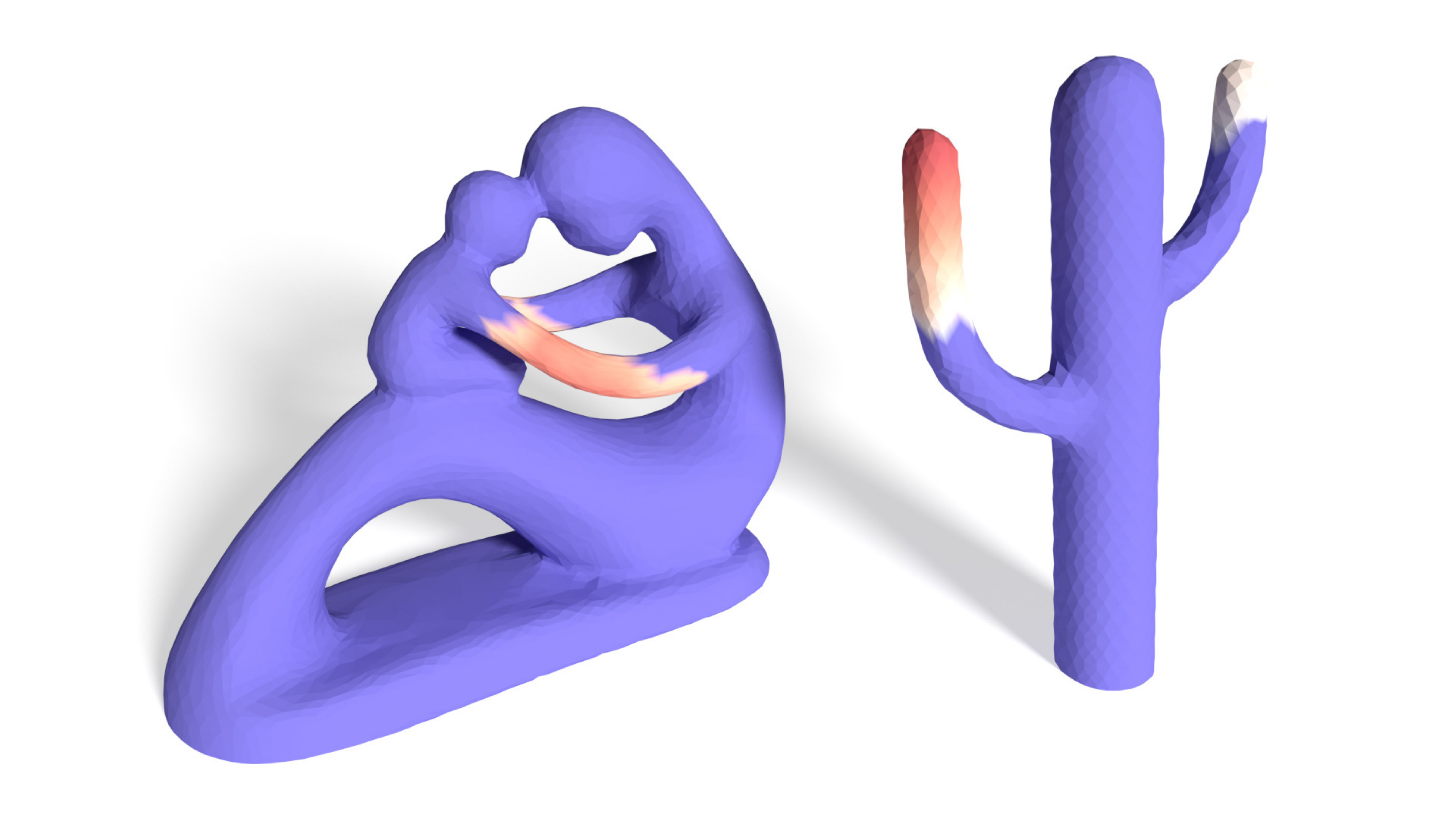}
\caption{Force regions computed for two fully solid models. We take connected components in high criticality regions as our FRs (red-white areas).}
\label{fig:FRs}
\end{figure}

The accuracy of the criticality map depends on the number of training samples $l$. While a large number of samples increases accuracy, the computational cost also increases proportionally. Figure~\ref{fig:ActualVSEstimateCriticalityMap} illustrates the criticality maps obtained with different number of training samples. We observed that using $5\%$ of the nodes in $\mathcal{S}_L$ for training produces an acceptable approximation of the criticality map such that in all of our examples, after the criticality values are estimated, top $10\%$ of the nodes with the largest criticality values always contain the ground truth critical instant. Figure~\ref{fig:FRs} illustrates the force regions we obtained for two different models. 

\paragraph{Weak Regions}

For a structure with material distribution $\boldsymbol{\rho}$, we need to find the maximum stress produced by each force instant to determine $\sigma_{cr}$ in \eqref{Eq:sigmaCritical} and solve \eqref{Eq:optimizationProblem}.
However, failure often occurs at certain regions of the object leaving the remainder safe at all times.
WRs help us constrain the regions of the structure where we seek the maximum stress $\mathcal{V}_{wr} \subset \mathcal{V}$.

In our algorithm, we use an approach similar to \cite{zhou2013weak} to determine the possible failure locations as our weak regions.
We determine WRs using modal analysis that involves solving the generalized eigenvalue problem

\begin{equation}
\lambda_j \boldsymbol{M}_g(\boldsymbol{\rho}) \boldsymbol{u}_j = -\boldsymbol{K}(\boldsymbol{\rho}) \boldsymbol{u}_j, \quad j=1,2,\ldots,m
\label{Eq:modalAnalysis}
\end{equation}  

where $\boldsymbol{u}_j$ is $j$'th eigenmode, $\lambda_j$ is the corresponding eigenvalue and $\boldsymbol{M}_g(\boldsymbol{\rho})$ is the mass matrix for the tetrahedral mesh.
Note that WRs are structure dependent, hence need to be updated at each step of the optimization.
To reduce computational cost, we use a lumped mass matrix and distribute each element's mass to its nodes equally, thus creating a sparse diagonal matrix $\boldsymbol{M}_g~\in~\mathbb{R}^{3n \times 3n}$. 

\begin{figure}
\centering
\includegraphics[width = 0.9\columnwidth]{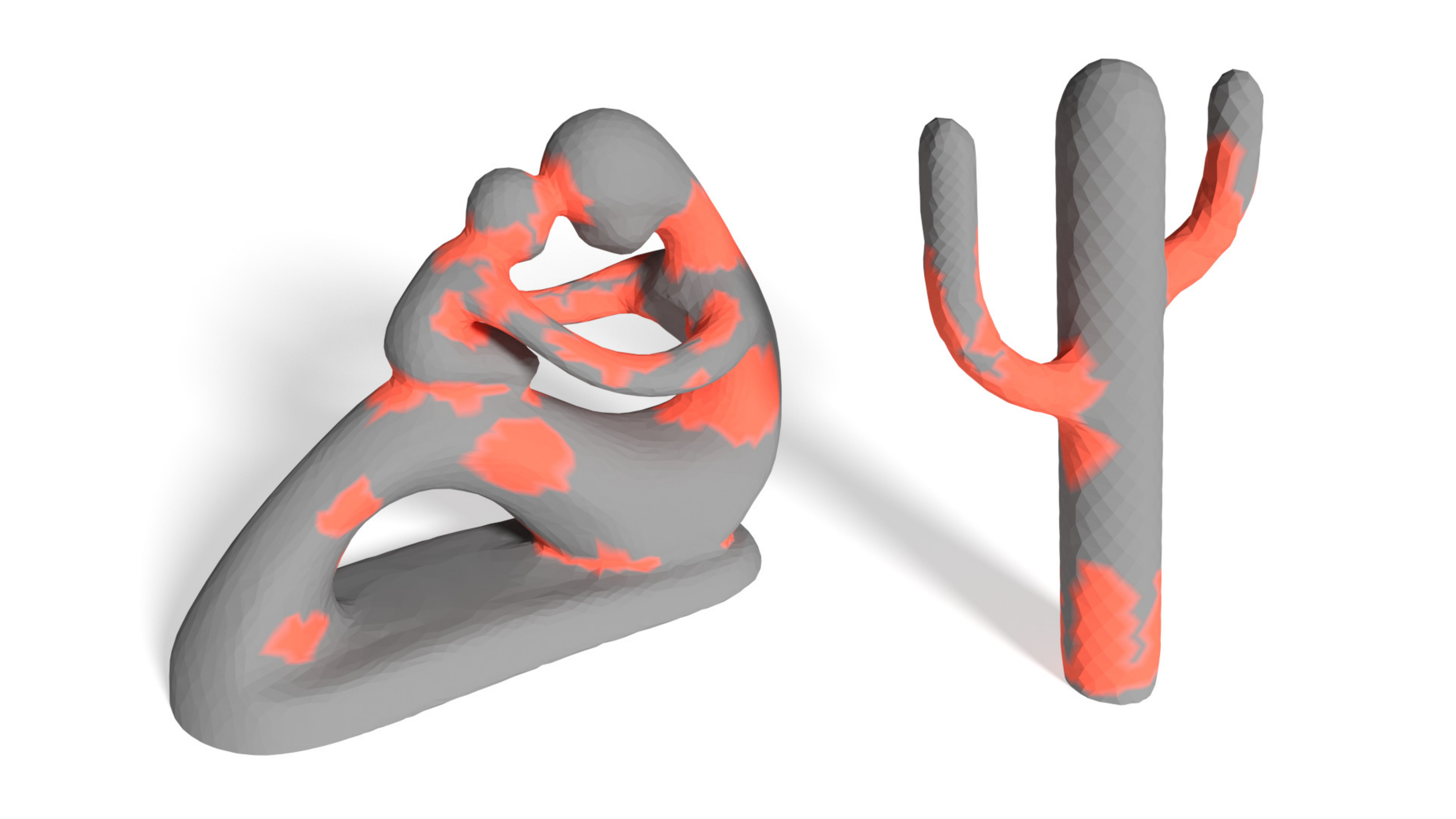}
\caption{Weak-regions are found around thin parts or sharp edges where stress concentration is likely to occur. Both are fully solid models and are fixed at the bottom.}
\label{fig:WRs}
\end{figure}

WRs can be extracted by computing the low frequency eigenmodes in \eqref{Eq:modalAnalysis} and identifying the nodes that experience large stresses under these deformations. In our examples, we use the first $15$ vibration modes and $2.5\%$ of the most stressed nodes to form our WRs. Different from \cite{zhou2013weak}, we combine all unique nodes obtained from modal analysis to construct the WRs. Figure \ref{fig:WRs} shows the WRs we obtained for two fully solid models ($\rho_e = 1 \enskip \forall e$). Note that WRs are found around possible stress concentration points such as thin parts and crease edges.

\paragraph{Critical Instant}

We solve \eqref{Eq:sigmaCritical} for $\sigma_{cr}$ in a much smaller domain defined by the FRs $\mathcal{S}_{fr} \subset \mathcal{S}_{L}$ and WRs $\mathcal{V}_{wr} \subset \mathcal{V}$. In particular, only the force instants captures with FRs are used, and maximum stresses are only sought in WRs.

We use a simple greedy hierarchical search to find the force instant creating the largest stress in the structure. For each FR island, we partition it into four segments and perform FEA by applying the force to their central nodes. We then further select and partition the segment that produces the highest stress within WRs, and repeat this process until converging to a single node. After repeating this process for all FR islands, we choose the maximum stress across all FR islands as $\sigma_{cr}$ for that particular optimization step. The stiffness matrix $\boldsymbol{K}$ and $\boldsymbol{C}_g$ need to be computed only once during these evaluations as the structure remains unchanged.  We thus factorize $\boldsymbol{K}$ once for each optimization step, and determine displacements and stresses using efficient forward and backward substitutions.

\subsection{Stress Singularity}
\label{sec:Stress Singularity}

Stress constrained mass minimization problems are prone to singularity issues \cite{sved1968structural,kirsch1990on,lee2012stress}.
For instance, in  \eqref{Eq:optimizationProblem}, the global optimum is obtained when $\boldsymbol{\rho}=\boldsymbol{0}$, making all elements void. To mitigate this problem, we establish a layer of fully solid boundary shell that is excluded from optimization. This approach overcomes the singularity problem by coercing the  optimization to employ material in the remaining inner volume as a way relieve the high stresses generated on the boundary shell. Enforcing a boundary shell also preserves the original outer surface.

\begin{figure}
\centering
\includegraphics[width = \columnwidth]{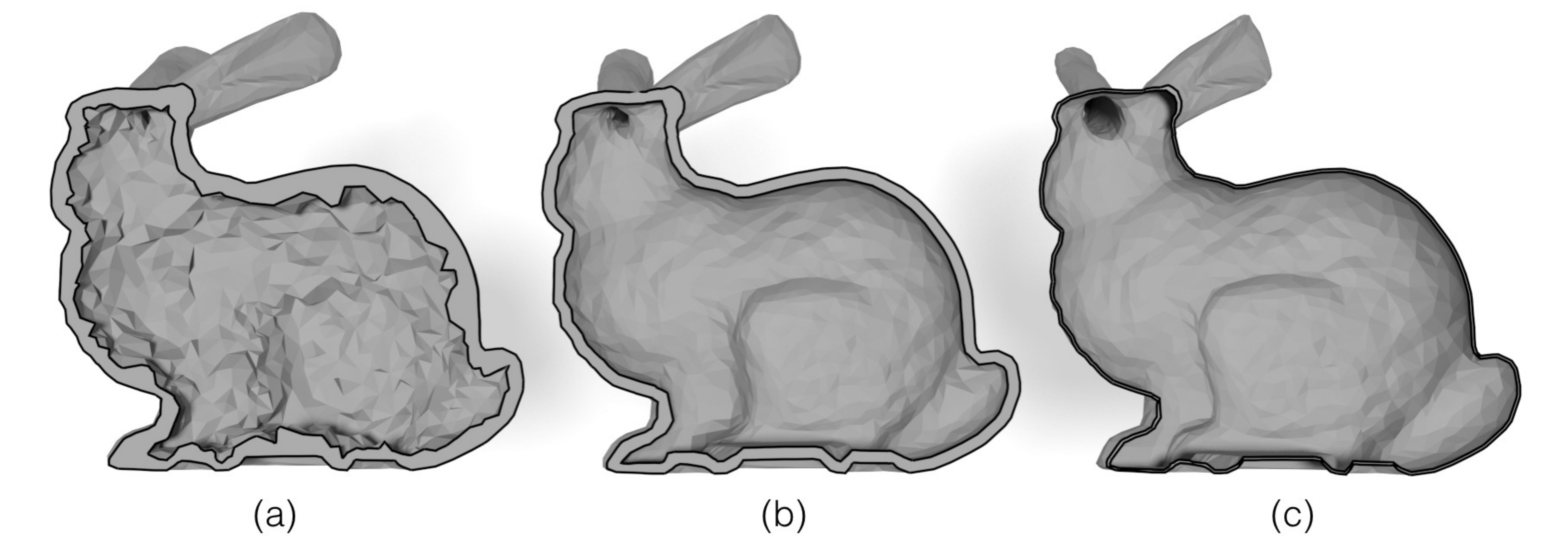}
\caption{To prevent undulations in the shell (a), we use offset surfaces to create a smooth boundary shell with prescribed thicknesses (b-c). In all three cases, the volumetric mesh has $29k$ elements.}
\label{fig:offsetSurfaces}
\end{figure}

All elements that contribute one or more vertices to the outer boundary $\mathcal{S}_0$ could serve as the boundary elements. For an arbitrary volumetric mesh, however, forming a solid shell using only these elements may introduce stress concentrations (Figure~\ref{fig:offsetSurfaces}(a)). To create a smooth and uniform thickness shell, we first generate an inner offset surface $\mathcal{S}_i$ using the method presented in \cite{musialski2015reduced}.
Then, we tetrahedralize the entire domain, which results in a smooth shell layer $\mathcal{V}_s \subset \mathcal{V}$ sandwiched by $\mathcal{S}_i$ and $\mathcal{S}_0$  (Figure~\ref{fig:offsetSurfaces}(b-c)). We use a uniform shell thickness prescribed by the user, which can be adjusted based on a 3D printer's minimum print thickness. Although we do not optimize the shell thickness, we discuss its effects in Section~\ref{sec:resultsAndDiscussion}.

\section{Model Reduction}
\label{sec:ModelReduction}

\begin{figure}
\centering
\includegraphics[width = \columnwidth]{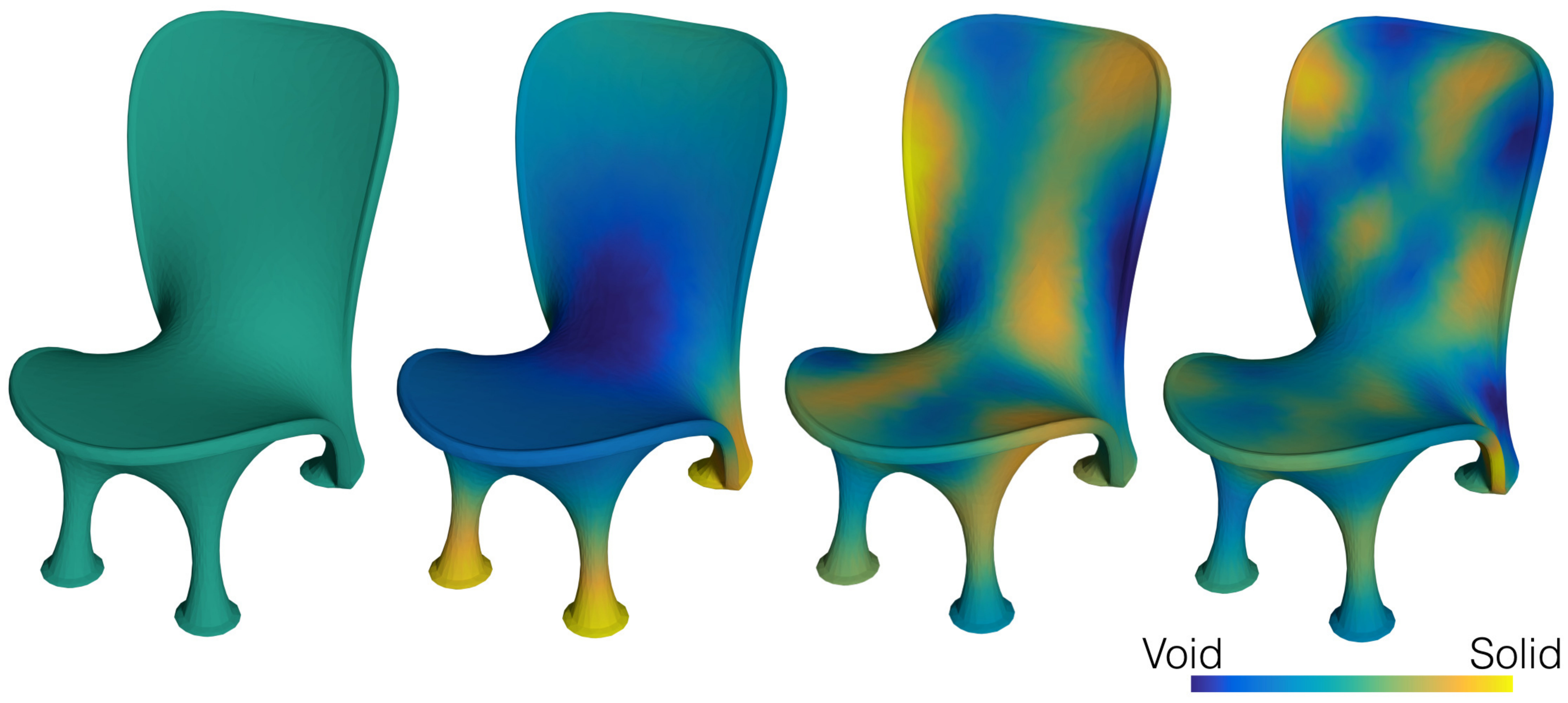}
\caption{Left to right: material distributions corresponding to $1st$, $5th$, $25th$ and $50th$ modes. Lower modes control global material distribution while higher modes enable local details.}
\label{fig:materialModes}
\end{figure}

The optimization problem \eqref{Eq:optimizationProblem} is typically very high-dimensional as the number of design variables is equal to the number elements in the volumetric mesh $\mathcal{V}$. This, in turn, has a significant impact on the computational performance. To accelerate optimization, we compute a set of material modes \cite{xu2015interactive}, which helps control the material distribution using only a small number of variables.  Material modes can be computed as the eigenvectors of the element-based graph Laplacian $\boldsymbol{\mathcal{L}} \in \mathbb{R}^{m \times m}$ defined on $\mathcal{V}$ by solving the generalized eigenvalue problem

\begin{equation}
\mu_j \boldsymbol{V} \boldsymbol{\gamma}_j = -\boldsymbol{\mathcal{L}} \boldsymbol{\gamma}_j, \quad j=1,2,\ldots,m
\label{Eq:materialModes}
\end{equation}

where $\mu_j$ are non-negative eigenvalues, $\boldsymbol{\gamma}_j$ are corresponding eigenvectors and $\boldsymbol{V} \in \mathbb{R}^{m \times m}$ is a diagonal matrix composed of $V_e$s.
Eigenvectors $\boldsymbol{\gamma}_j$ are orthogonal and smooth scalar functions that spectrally decompose the material distribution \cite{zhou2005large,zhang2010spectral}. The first  mode represents the homogeneous material distribution and while the level of detail control increases with higher frequencies (Figure~\ref{fig:materialModes}).
We assemble the first $k$ eigenvectors to form the reduced order basis  $\boldsymbol{\Gamma} = [\boldsymbol{\gamma}_1, \boldsymbol{\gamma}_2, \ldots \boldsymbol{\gamma}_k]$ so that the material distribution can be written as

\begin{equation}
\boldsymbol{\rho} = \boldsymbol{1} + \boldsymbol{\Gamma} \boldsymbol{\alpha},
\label{Eq:reducedOrderDensity}
\end{equation}

where $\boldsymbol{\alpha} = [\alpha_1, \alpha_2, \ldots \alpha_k]^T$ is the design vector for the reduced order problem.
This formulation allows us to trivially enforce fully solid material on the boundary shell elements by by setting the corresponding rows in the reduced  basis matrix $\boldsymbol{\Gamma}_e$ to be $\boldsymbol{0}$. This way the entries in $\boldsymbol{\alpha}$ can take on any value during the optimization without violating the geometrical constraints. 

\paragraph{Logistic Function} 
In our reduced order formulation, we use a logistic function $G(x)$ to penalize the intermediate values of $\rho_e$ by modifying \eqref{Eq:reducedOrderDensity} as

\begin{subequations}
\label{Eq:logistic}
\begin{align}
\boldsymbol{\rho} &= G(\boldsymbol{\Gamma} \boldsymbol{\alpha}) \label{Eq:logisticDensity}\\
G(x) &= 1/[1+e^{(\kappa(x-x_0))}]. \label{Eq:logisticFunction}
\end{align}
\end{subequations}

Here, $\kappa$ and $x_0$ determine the steepness and inflection point of the logistic function.
Note that $x_0$ should be adjusted to satisfy $G(\boldsymbol{0}) \approx \boldsymbol{1}$ to ensure that the elements on the boundary shell are solid.
While increasing $\kappa$  intensifies binarization by pushing the intermediate densities  toward $0$ and $1$, it also hampers convergence. We use $\kappa=5$ for all of our examples.   

In addition to binarizing the intermediate densities, the use of logistic function in \eqref{Eq:logisticDensity} guarantees $\rho_e \in [0,1]~\forall e \in \mathcal{V}$ for $-\infty<\boldsymbol{\alpha}<\infty$.
This allows us to remove a large number of constraints $\boldsymbol{0} \leq \boldsymbol{\rho} \leq \boldsymbol{1}$ from our reduced order optimization problem.

\subsection{Reduced Order Problem}
Applying \eqref{Eq:logistic} to our optimization formulation \eqref{Eq:optimizationProblem}, the reduced order optimization problem can be stated as

\begin{equation}
\begin{aligned}
& \underset{\boldsymbol{\alpha}}{\text{minimize}}
& & M(\boldsymbol{\alpha})  = G(\boldsymbol{\Gamma} \boldsymbol{\alpha}) \cdot \boldsymbol{V} \\
& \text{subject to}
& & \boldsymbol{K}(\boldsymbol{\alpha}) \boldsymbol{u}_i = \boldsymbol{f}_i \quad \forall i \in \mathcal{S}_{fr}, \\
& & & \sigma_{cr}(\boldsymbol{\alpha}) \leq \sigma_y,\\
\end{aligned}
\label{Eq:reducedOptimizationProblem}
\end{equation}
where 
\begin{equation}
\sigma_{cr} = \underset{\boldsymbol{i}}{\text{max}}( \underset{\boldsymbol{e}}{\text{max}}( \enskip \sigma_e^{vm} \enskip)) \quad \forall i \in \mathcal{S}_{fr} \enskip \text{and} \enskip \forall e \in \mathcal{V}_{wr},
\label{Eq:reducedSigmaCritical}
\end{equation}
for $\boldsymbol{\Gamma}_e = \boldsymbol{0} \quad \forall e \in \mathcal{V}_s$.

A benefit of the reduced order formulation is that the number of new design variables $k$ can be markedly small compared to the number of original variables $m$ in \eqref{Eq:optimizationProblem}. This number is independent of the input mesh and needs to be prescribed by the user. 
Because the structural optimization algorithm involves a large number of costly FEA evaluations per iteration, we found $k \leq 15$ to provide a favorable tradeoff between speed and expressiveness. Hence, we use $15$ material modes in all of our examples, unless otherwise stated. The optimization starts with a fully solid model ($\boldsymbol{\alpha} = \boldsymbol{0}$). Our approach is predicated on the assumption that this starting solution is feasible, hence amenable to lightweighting through optimization.

At the end of the optimization, the resulting material distribution may still contain elements with intermediate densities. In such cases, we  threshold the gray scale material distribution  to obtain a fully binarized solution. In our examples, we use $\rho=0.5$ as the threshold; elements with lower densities are set to void. Nonetheless, there exists more sophisticated thresholding methods \cite{hsu2005interpreting}. Another positive byproduct of the reduced order approach is that the resulting material distribution after binarization typically does not suffer from a checkerboard effect, as the reduction leads to smooth material modes in $\mathcal{V}$, especially for $k \ll m$. This, in turn, helps alleviate exhaustive post-processing.

Note that the material modes are precomputed as they depend only on $\mathcal{V}$. Because only the first $k$ modes are used in the reduced order formulation, they can be computed efficiently using iterative methods. We use ARPACK for this purpose \cite{lehoucq1998arpack}.
In order to solve \eqref{Eq:reducedOptimizationProblem}, we use sequential quadratic programming \cite{nocedal2006numerical}.
\myColor{Our code solves linear systems using the Eigen library's SimplicialLDLT sparse solver \cite{eigenweb}.}

Our hierarchical search method and density based shape representation allow us to compute gradients analytically. We compute gradients of the critical stress with respect to the design variables using the adjoint method \cite{Paris2010stress}. We use p-norm approximations ($p=15$) for the max functions. \myColor{Details are given in Appendix~\ref{sec:Appendix}.}

\begin{figure}
\centering
\includegraphics[width = \columnwidth]{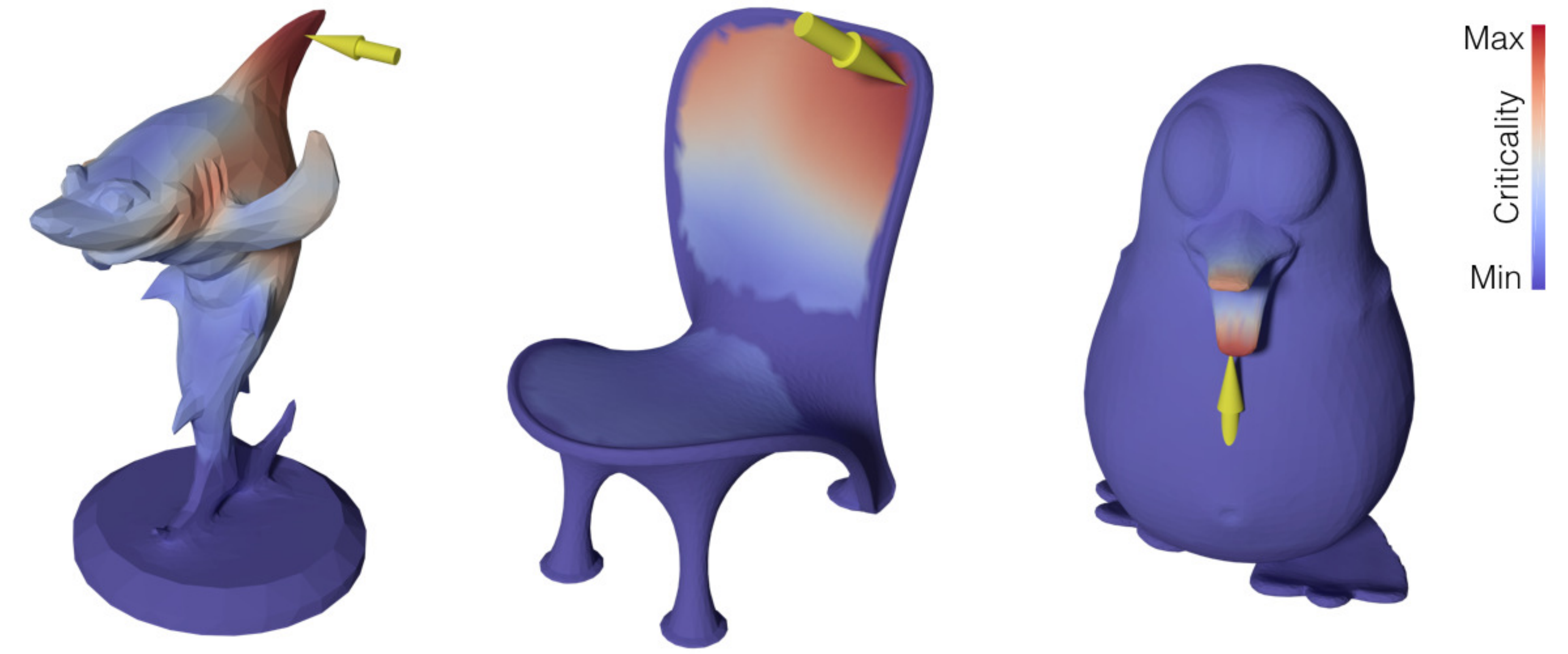}
\caption{Critical force instants are shown on estimated criticality maps. All models are fully solid and are fixed at the bottom. }
\label{fig:Analysis}
\end{figure}

\section{Results and Discussion}
\label{sec:resultsAndDiscussion}
   
\subsection{Criticality Analysis}

\begin{figure}
\centering
\includegraphics[width = \columnwidth]{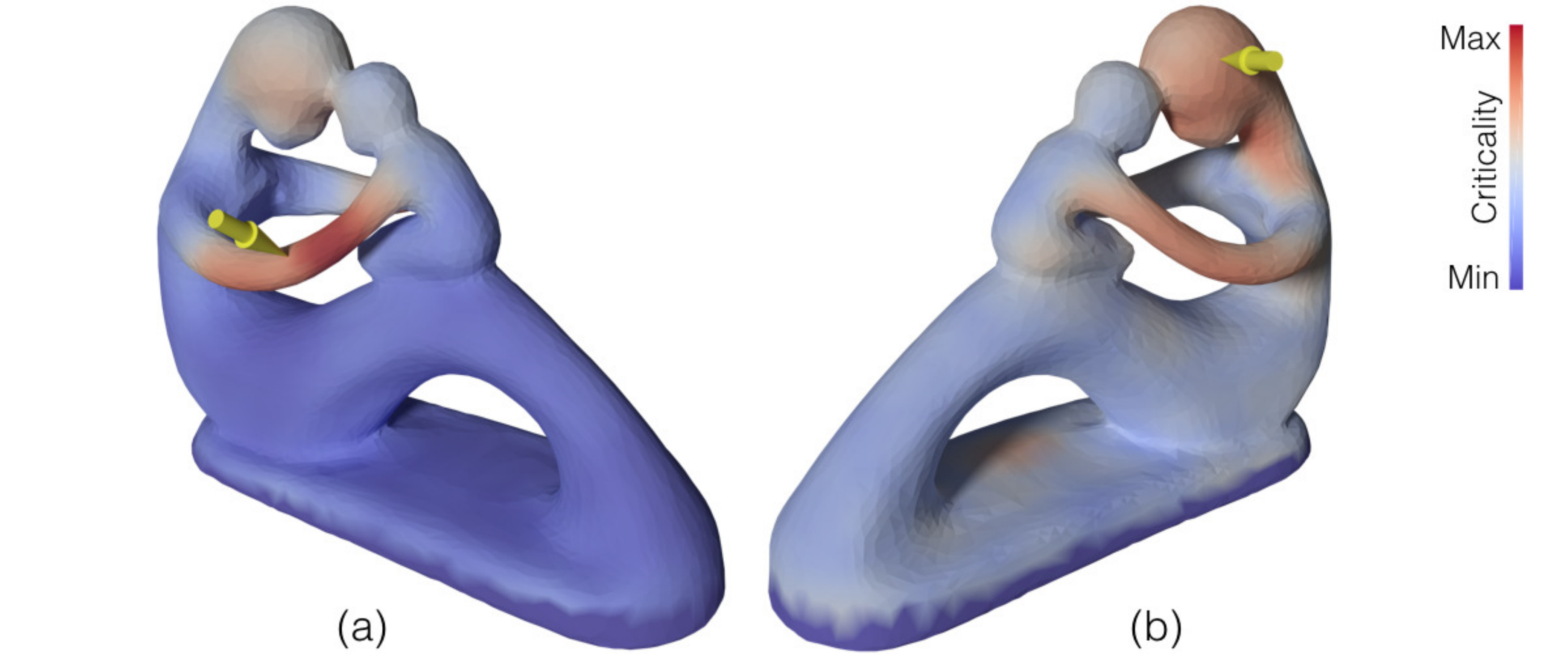}
\caption{Critical instant can change significantly during optimization. While forces around the arms are critical for the fully solid model (a), the critical instant shifts to the mother's temple as the structure is hollowed (b).}
\label{fig:AnalysisFertility}
\end{figure}

Figure~\ref{fig:Analysis} illustrates the results of our critical instant analysis  on a collection of fully solid models.
In all cases presented in this paper, our approach involving criticality map estimation followed by a hierarchical search is able to determine the true critical force instant (yellow arrows). We verified this match using an expensive brute force search method. As shown in Figure~\ref{fig:Analysis}, our estimated criticality map captures the true critical instant quite well in that it finds the most critical force to be at the point with the highest estimated criticality (dark red). Nonetheless, our analysis can tolerate inaccuracies in the estimated criticality map. Figure~\ref{fig:AnalysisFertility}(a) shows such a case. Although the criticality map estimates the most critical point to be on the arm closer to the baby (dark red), our algorithm subsequently finds the true critical instant in that vicinity by searching an expanded area forming the force region.

\begin{figure}
\centering
\includegraphics[width = \columnwidth]{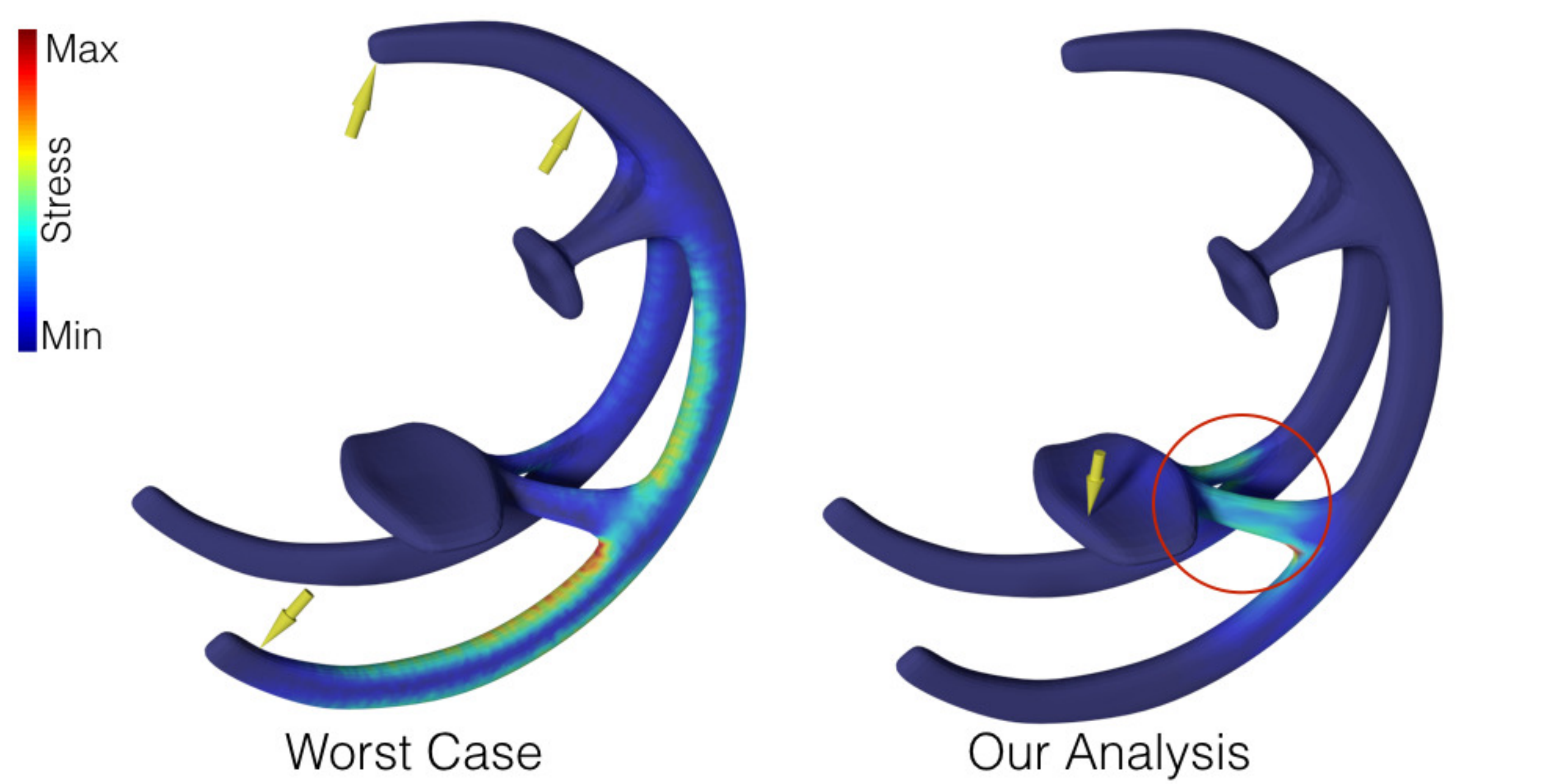}
\caption{A comparison between worst case structural analysis \protect\cite{zhou2013weak} and our method. 
While worst case analysis predicts the critical forces well for handling scenarios, it may miss contextually relevant critical instants during actual use cases. Our analysis, with delineated contact regions and boundary conditions,  predicts the critical instants to be around the seat of the rocker.}
\label{fig:WorstCaseVSOurs}
\end{figure}

Figure~\ref{fig:AnalysisFertility} illustrates the critical instants for two different material distributions.
Because we construct the criticality map for each step of the optimization, the force regions are updated and the change in the critical instant is captured  well. In all of our examples, the critical force instant is always contained in the identified force regions. 

\paragraph{Comparison} Figure~\ref{fig:WorstCaseVSOurs} compares our critical instant analysis with the worst-case structural analysis of Zhou \textit{et. al.}~\shortcite{zhou2013weak}.
The worst-case structural analysis method is designed to predict the critical force configurations that develop during handling an object.
Thus, it may have a tendency to overpredict the weakness, resulting in overengineered solutions if used for structural optimization.
In our method, because we delineate the contact regions and the displacement constraints to reflect the knowledge of actual use, our approach captures the critical forces that are more likely to be encountered during a product's nominal use. For instance, the region encircled in Figure~\ref{fig:WorstCaseVSOurs} for our analysis shows the high stress region worst-case analysis fails to capture. The boundary conditions and the contact regions used in our analysis are shown in Figure~\ref{fig:Results}.

\subsection{Structural Optimization}

\begin{figure}
\centering
\includegraphics[width = \columnwidth]{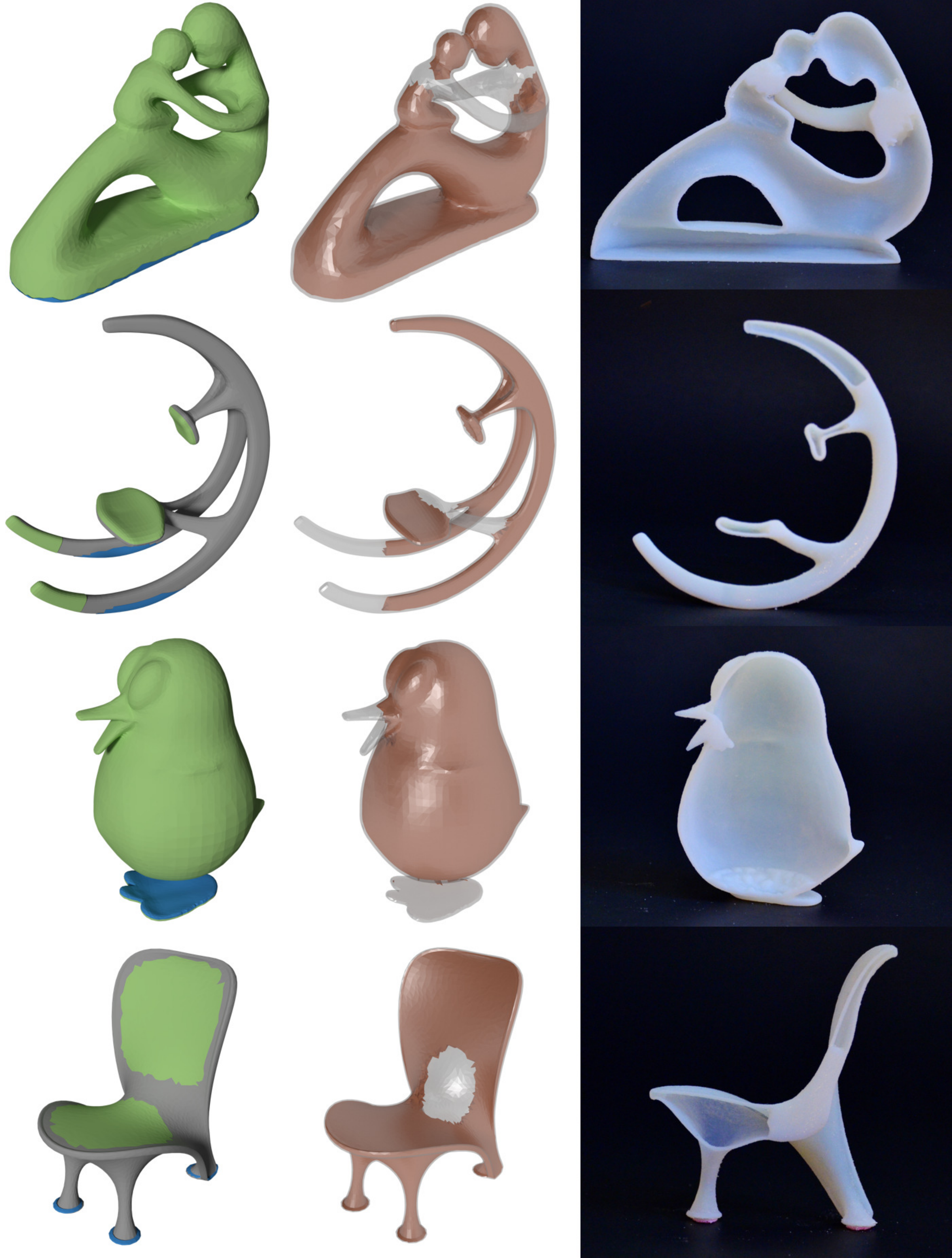}
\caption{Structural optimization results. Left-to-right, problem setups with fixed boundary conditions (blue) and contact regions (green), our optimized structures and 3D printed cut-outs of optimal results. Red shows the removed material.}
\label{fig:Results}
\end{figure}

Figure~\ref{fig:Results} illustrates the results of our method on various 3D models. The displacement-constrained regions are shown in blue. The contact regions are shown in green. Our reduced order  optimization  detects the parts of the objects where high stresses may develop and distributes material accordingly. For objects with small weak regions such as the fertility and penguin models, our optimization allows us to preserve the same structural strength that their fully solid versions possess, while shedding a large portion of the mass.
Similarly, our algorithm performs well for models with thin elements such as the chair by utilizing mass around the stress carrying regions of the object.
Notice the arms supporting the seat in the rocking chair and the back support in the chair model.

In these examples, we  achieved $50\%$ to $90\%$ mass reduction. Table~\ref{tab:Results} summarizes the weight reduction together with various other metrics relevant to these models.

\begin{figure}
\centering
\includegraphics[width = \columnwidth]{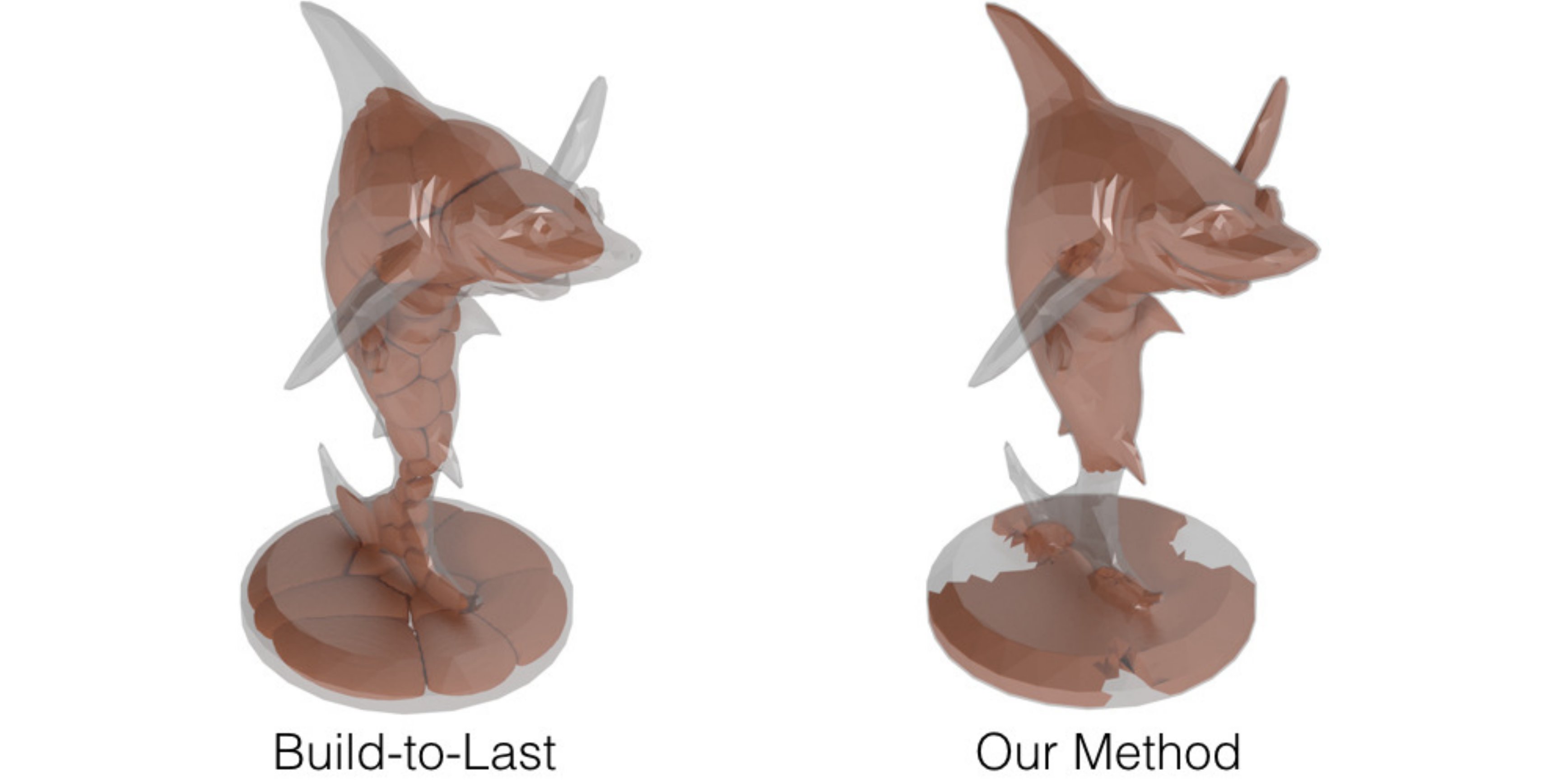}
\caption{A comparison between build-to-last \protect\cite{lu2014build} and our method. Our optimization approach produces a lighter structure while sustaining any possible force applied on the boundary. Build-to-last optimizes the structure for  a single static force.}
\label{fig:BuildToLastVSOurs}
\end{figure}

\paragraph{Comparison}
In Figure~\ref{fig:BuildToLastVSOurs}, we compare our approach with the build-to-last method  \cite{lu2014build}.
We impose a force budget of $20N$ applied anywhere on the surface of the shark. Our optimum result weighs $33\%$ less than the build-to-last structure that takes a prescribed force location as input\footnote{In \cite{lu2014build}, we could not identify the force magnitude being used.}.
Unlike build-to-last, our algorithm hollows the fins and the nose where high stresses cannot be generated in any force configuration.
Also, our method generates a three-pronged rib structure at the base, possibly to accommodate the forces that can be applied laterally in all arbitrary directions. 

\begin{figure*}
\centering
\includegraphics[width = \textwidth]{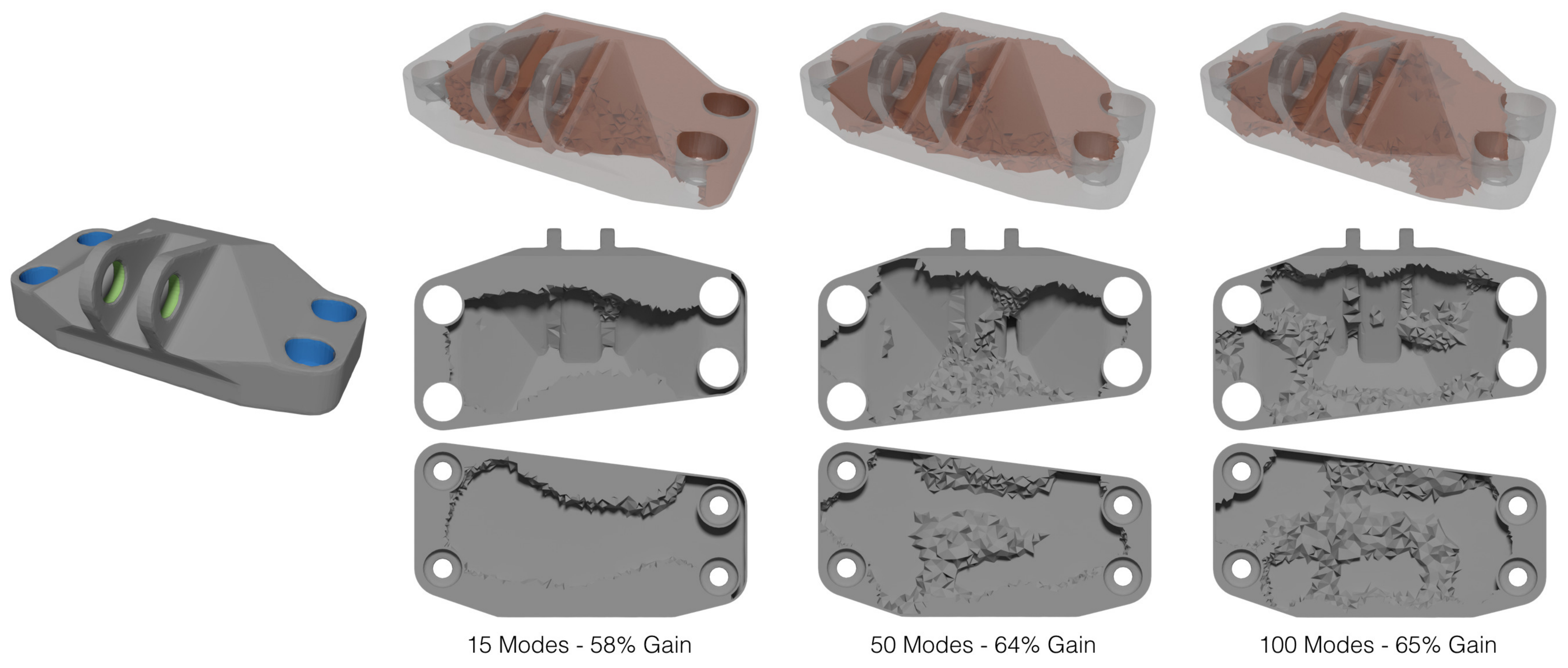}
\caption{Bracket model (left) is optimized using three different number of material modes. As the number of material modes increases, optimization can perform more localized alterations resulting in smaller mass structures at a cost of higher computational complexity.}
\label{fig:GEBracketDifferentModes}
\end{figure*}

\paragraph{Number of Material Modes} Figure~\ref{fig:GEBracketDifferentModes} illustrates the effect of the number of material modes.
We optimized the bracket using $15$, $50$ and $100$  modes for the same force budget and boundary conditions.
As shown, higher material modes allow for a finer local shape control. Hence, a larger mass reduction is obtained with increasing number of material modes (see Table~\ref{tab:Results}). Although larger number of modes allow our algorithm to remove more material through local control, computational  cost also increases. In Table~\ref{tab:Results}, per iteration computation times are given for the bracket model using different number of material modes. As the number of modes increases, computational cost increases significantly, while only a minor improvement in further mass reduction is achieved.  

\begin{figure}
\centering
\includegraphics[width = \columnwidth]{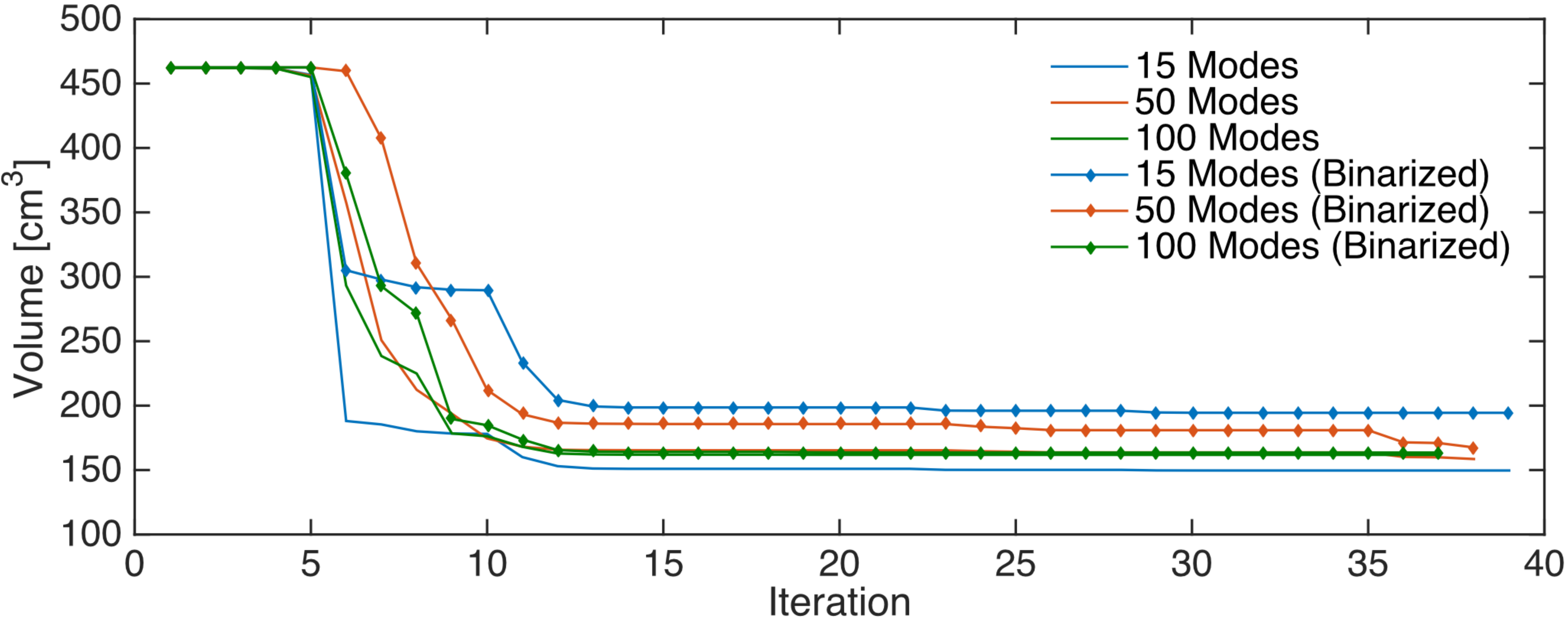}
\caption{Convergence plot for the bracket model using three different material modes. Binarized values are also shown.}
\label{fig:Convergence}
\end{figure}

\paragraph{Convergence} Figure~\ref{fig:Convergence} shows the convergence profiles for the  bracket model. While convergence is achieved after a similar number of iterations, the smaller number of material modes tend to leave more intermediate density elements in the optimized distribution, resulting in a larger mass when binarized.

\begin{figure}
\centering
\includegraphics[width = \columnwidth]{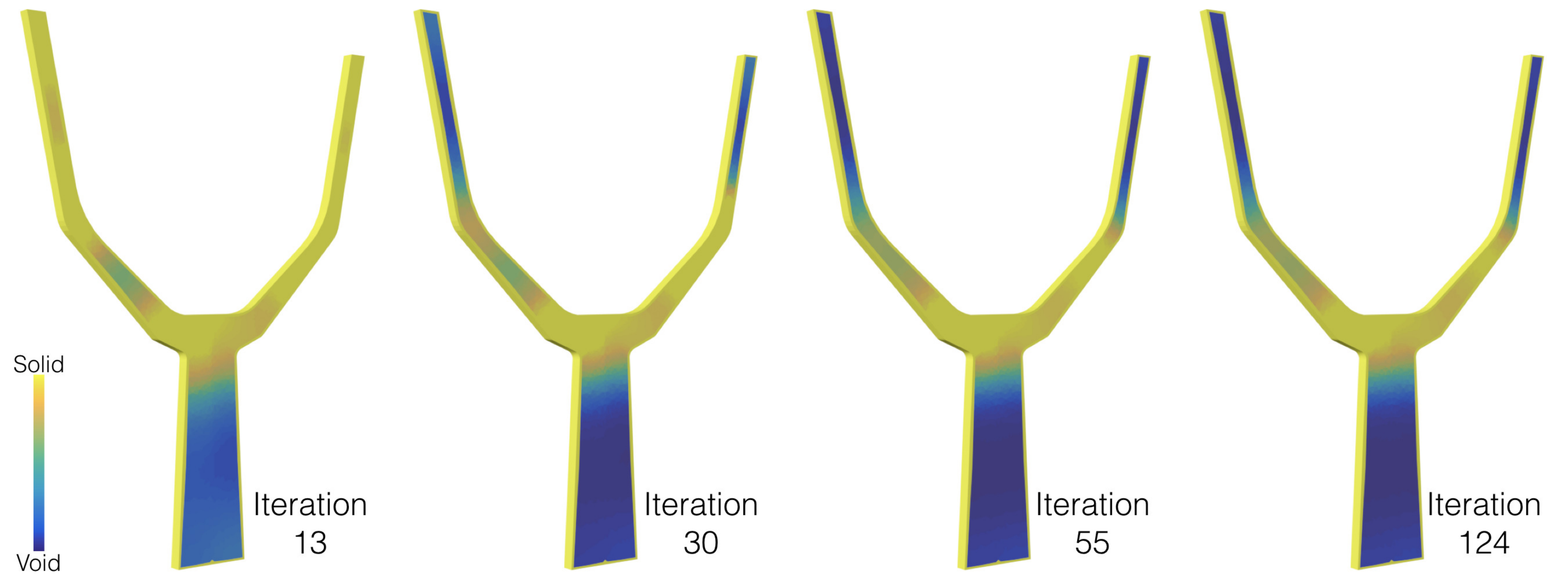}
\caption{\myColor{Evolution of material distribution throughout the optimization process for a symmetrical slingshot model. Optimization is initialized with fully solid object. Left to right  material distribution converges to the resulting state. The model is fixed at the bottom.}}
\label{fig:ConvergenceSymmetric}
\end{figure}

\myColor{One might worry that in a symmetrical object, the critical point could jump from side to side, with every incremental improvement to one side causing a symmetric worsening of the other side, and convergence never being reached. We attempted to trigger this potential failure by creating a test case with a carefully-constructed perfectly symmetric boundary and tetrehedral mesh (Figure~\ref{fig:ConvergenceSymmetric}). Our algorithm nonetheless converged in $124$ steps. We conjecture that the homogenous (first) material mode tends to absorb enough of the change to avoid oscillation; though it could also be that low-level numerical asymmetries, \textit{e.g.}~ordering effects in linear algebra routines, account for the convergence.}

\paragraph{Boundary shell thickness}

\begin{figure}
\centering
\includegraphics[width = 0.9\columnwidth]{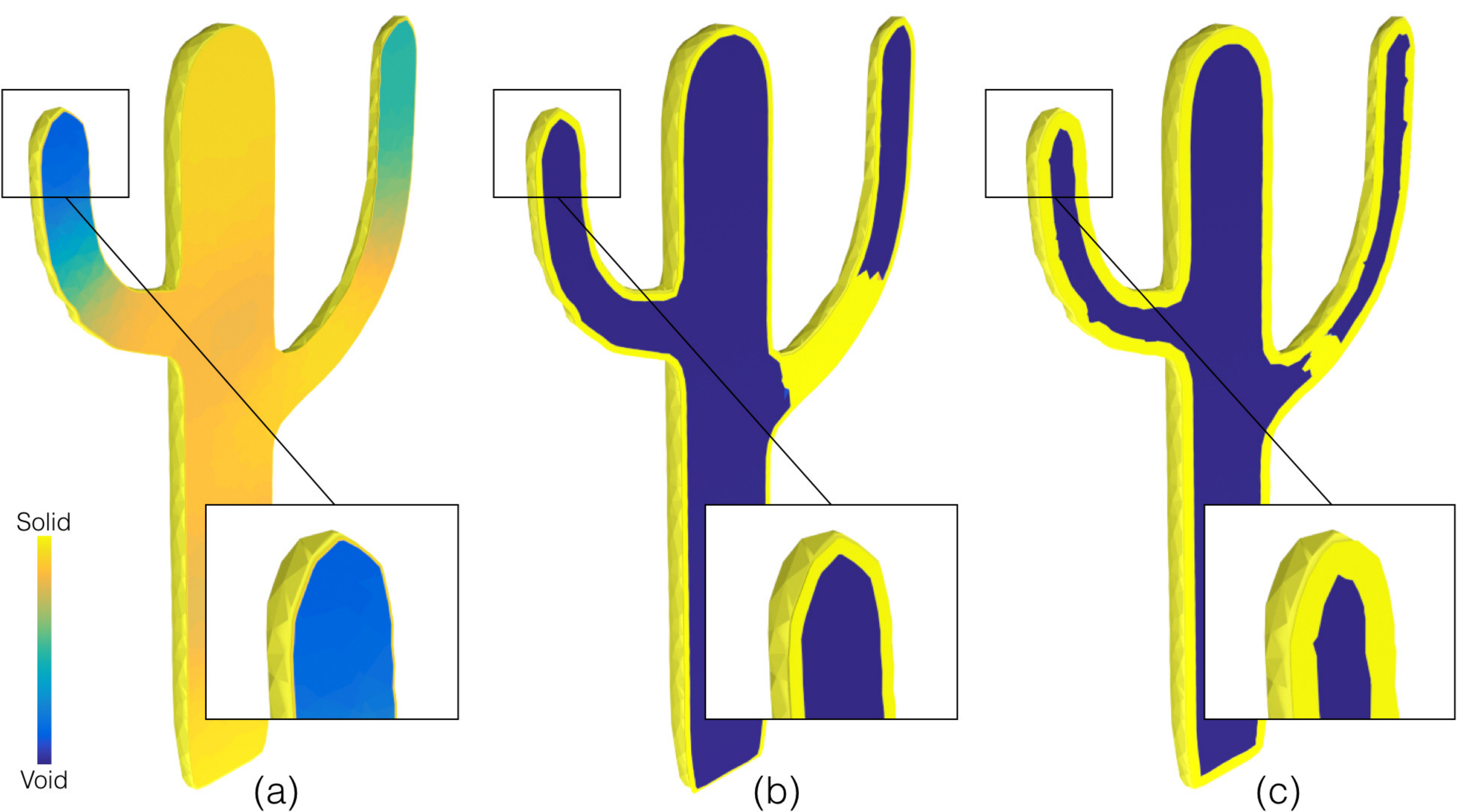}
\caption{Effect of boundary shell thickness on material distribution. From (a) to (c), shell thickness is increased while all other parameters are kept constant. For very small thickness values, our algorithm fails to converge to a binary material distribution using only few material modes due to high local compressive stresses. Close-up images show the shell thickness for each case.}
\label{fig:boundaryShellThickness}
\end{figure}

There is a trade-off between the shell thickness and the flexibility of our algorithm in generating an internal structure.
As the boundary shell gets thicker, its contribution to the structure's strength becomes more prominent, thereby shrinking the design space for our algorithm.
On the other hand, too thin boundary shells can lead to large local compressive stresses that renders optimization infeasible.
In such cases, our method fails to converge to a varied solid versus void material distribution (Figure~\ref{fig:boundaryShellThickness}(a)).
The main reason is that high local compressive stresses encourages the optimization to perform local thickening around the force application points. However, low frequency modes can alter the material distribution only in large chunks.
Therefore, fine-level local modifications cannot be achieved unless a large number of material modes are used.
Figure~\ref{fig:boundaryShellThickness} illustrates the effect of the shell thickness on the resulting material distribution.
With a proper choice of the thickness (Figure~\ref{fig:boundaryShellThickness}(b)), our method is able to reduce mass by $72\%$ compared to fully solid model, while the reduction was only $17\%$ and $54\%$ for (a) and (c), respectively.

\subsection{Validation and Performance}

\paragraph{Fabrication} We 3D printed our optimum results on an OBJET Connex printer using inkjet printing technology.
We use VeroWhitePlus material with a yield strength of $50 MPa$, a Young's modulus of $2.1 GPa$,  and  a Poisson's ratio of $0.3$  \cite{polyjetDatasheet}.

When soluble support material is used, the boundary shell can be pierced by small holes to empty the internal support material. We observed that our reduced order method has a tendency to create only a small number of inner void regions (especially for a small number of material modes) and thus the support material can be removed with minimal alterations. To avoid trapping support material, resulting models can be printed in several pieces and glued together after cleaning.

\begin{figure}
\centering
\includegraphics[width = \columnwidth]{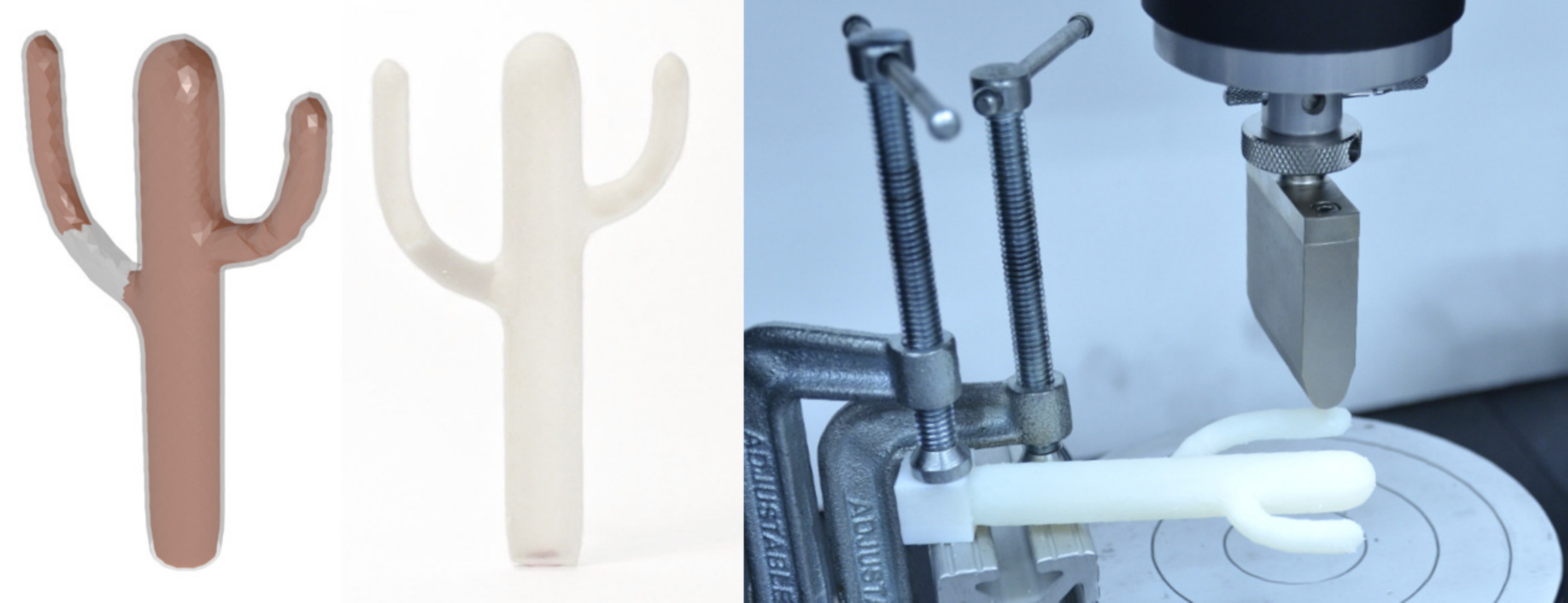}
\caption{Compression tests on the optimized cactus. Left and middle images show the resulting model and a 3D printed cutout.}
\label{fig:CactusPhysicalExperiment_1}
\end{figure}

\paragraph{Physical Tests} We performed compression tests on our optimized cactus model to physically evaluate the strength of the 3D printed models, as shown in Figure~\ref{fig:CactusPhysicalExperiment_1}. We used an INSTRON universal testing machine and ran compression tests on the optimized cactus model. For comparison, we chose an identically weighing uniform thickness cactus. We performed the same compression test on the long arm and measured the failure load. For our optimized model, we measured the failure force to be $36.82N$. The uniform thickness model snapped at $20\%$ less force of $29.69N$. 

\begin{figure}
\centering
\includegraphics[width = \columnwidth]{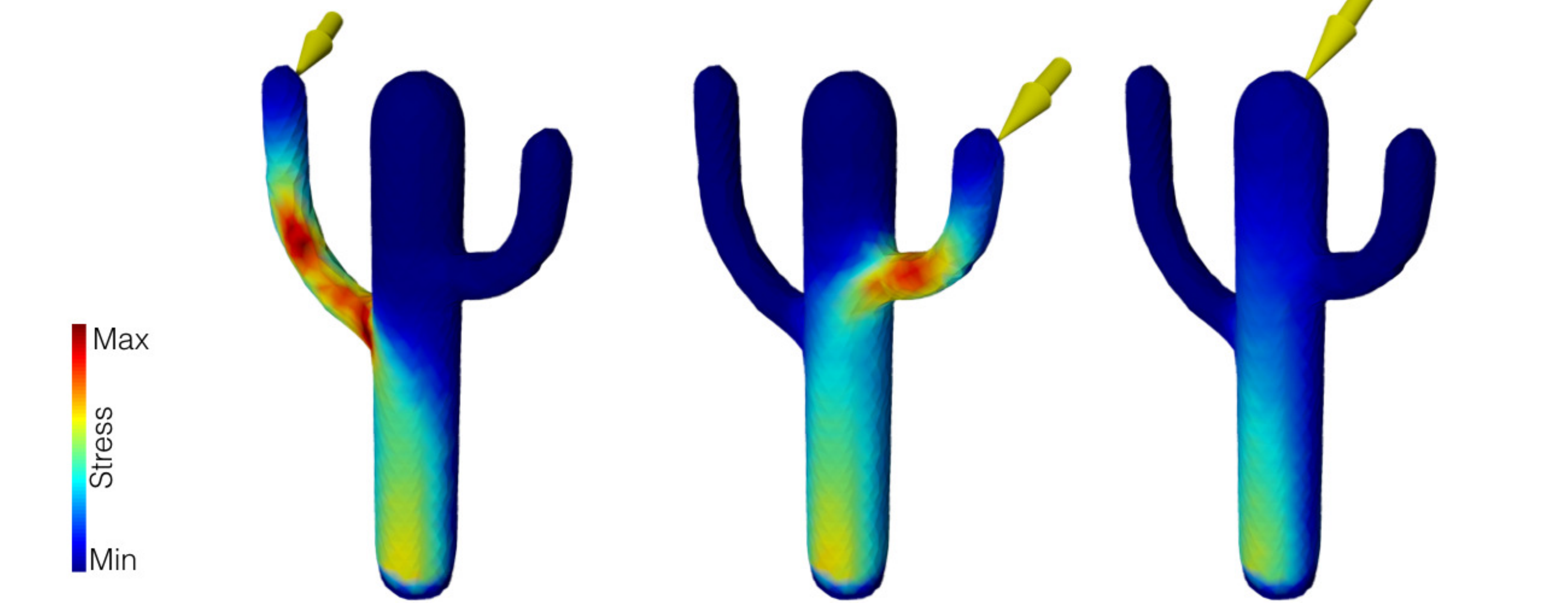}
\caption{Stress distributions when the same magnitude force is applied to three different locations on the optimized cactus. Analysis results match  the compression tests in that failure occurs on the long arm, short arm and trunk, in order. }
\label{fig:CactusPhysicalExperiment_2}
\end{figure}

For our optimized model, we measured the failure force of $39.62N$ for the short arm, and  $74.22N$ for the trunk.
The  test results agree well with our critical instant analysis in that the tip of the long arm turns out to be physically the most critical force point in the optimum model and there is no need to add material to either the short arm or the trunk.
Figure~\ref{fig:CactusPhysicalExperiment_2} illustrates the stress distribution when a $35N$ force is applied to the same three points as the physical tests. It can be observed that the short arm and the trunk are quite safe, while very high stresses are present on the long arm.

\begin{table*}
\small
\caption{\myColor{Performance of our algorithm on a variety of models. The number of FEAs and analysis times in columns 4 and 6 are average and given per step of critical instant analysis. Performance of the brute force approach is also shown for benchmarking. Columns 7 and 8 are number of iterations to fully optimize the structures and average times per iteration of our structural optimization, respectively.}} 
\centering 
\begin{tabular}{l ccccccccc} 
\hline\hline 
\multirow{2}{*}{Model} & \multirow{2}{*}{Elements} & \multicolumn{2}{c}{\# of FEA} & \multicolumn{2}{c}{Analysis Time [s]}  & \multirow{2}{*}{Iteration} & \multirow{2}{*}{Time[s]} &\multicolumn{2}{c}{Volume [cm\textsuperscript{3}]}\\ \cline{3-4} \cline{5-6} \cline{9-10}
& & Brute Force  & Our Method  & Brute Force & Our Method & & & Initial & Optimized\\ [0.5ex]
\hline 
Cactus     				& 25229		& 1658	& 216	& 17.69		& 4.35		& 49	&	39.61	& 68.990	& 18.800\\
Slingshot				& 54290		& 2192	& 148	& 89.02			& 10.66		& 124	&	72.52	& 17.822	& 9.451\\
Fertility    			& 57006		& 3914	& 248	& 131.85	& 20.68		& 122	&	83.33	& 54.023 	& 16.350\\
Test (Small Force)		& 58555		& 822	& 69	& 26.53		& 7.52		& 180	&	59.15	& 4.443		& 1.216\\
Test (Large Force)		& 58555		& 822	& 69	& 26.53		& 7.52		& 223	&	59.15	& 4.443		& 1.300\\
Chair   				& 58848		& 663	& 52	& 27.96		& 13.28		& 104	&	75.32	& 17.076	& 6.192\\
Penguin        		& 68615		& 3035	& 304	& 149.80	& 45.13		& 31	&	132.06	& 91.643 	& 10.211\\
Shark			 		& 70397		& 4282	& 273	& 178.93	& 18.10		& 72	&	115.07	& 66.454	& 14.997\\ 
Rocking Chair  		& 78025		& 1531	& 103	& 85.48		& 11.17		& 160	&	154.32	& 13.015	& 6.709\\ 
Bracket (k = 15)   	& 111498	& 408	& 55	& 70.85		& 48.87		& 39	&	301.01	& 462.520	& 194.434\\ 
Bracket (k = 50)  		& 111498	& 408	& 55	& 70.85		& 48.87		& 38	&	382.49	& 462.520	& 167.552\\ 
Bracket (k = 100) 		& 111498	& 408	& 55	& 70.85		& 48.87		& 37	&	487.44	& 462.520	& 163.581\\ 

\hline 
\end{tabular}
\label{tab:Results}
\end{table*}

\begin{figure}
\centering
\includegraphics[width = \columnwidth]{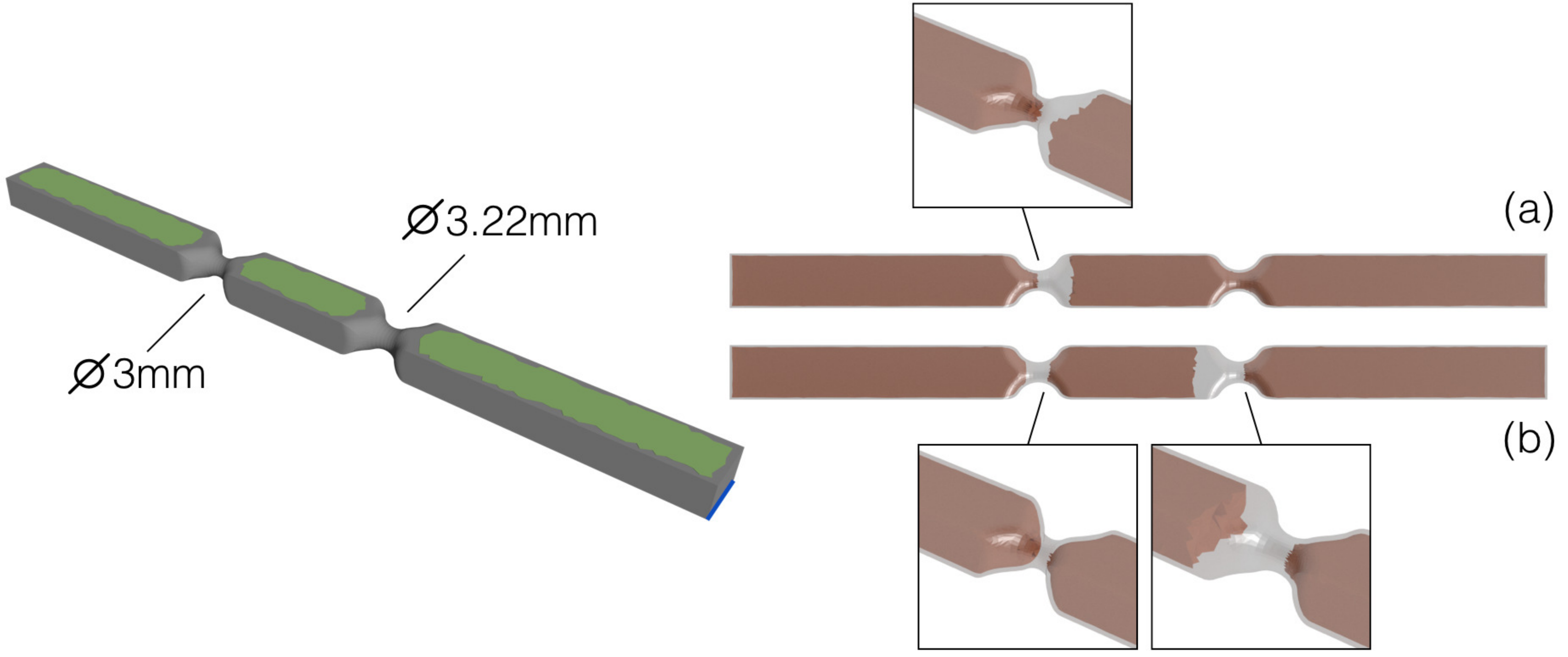}
\caption{A model with two failure-prone regions optimized for two different force budgets. For a force budget of $4N$ (a), material is only placed around the thinnest region while both critical regions are beefed up  for a larger force budget of $5.5N$ (b).}
\label{fig:TestExample}
\end{figure}

\begin{figure}
\centering
\includegraphics[width = \columnwidth]{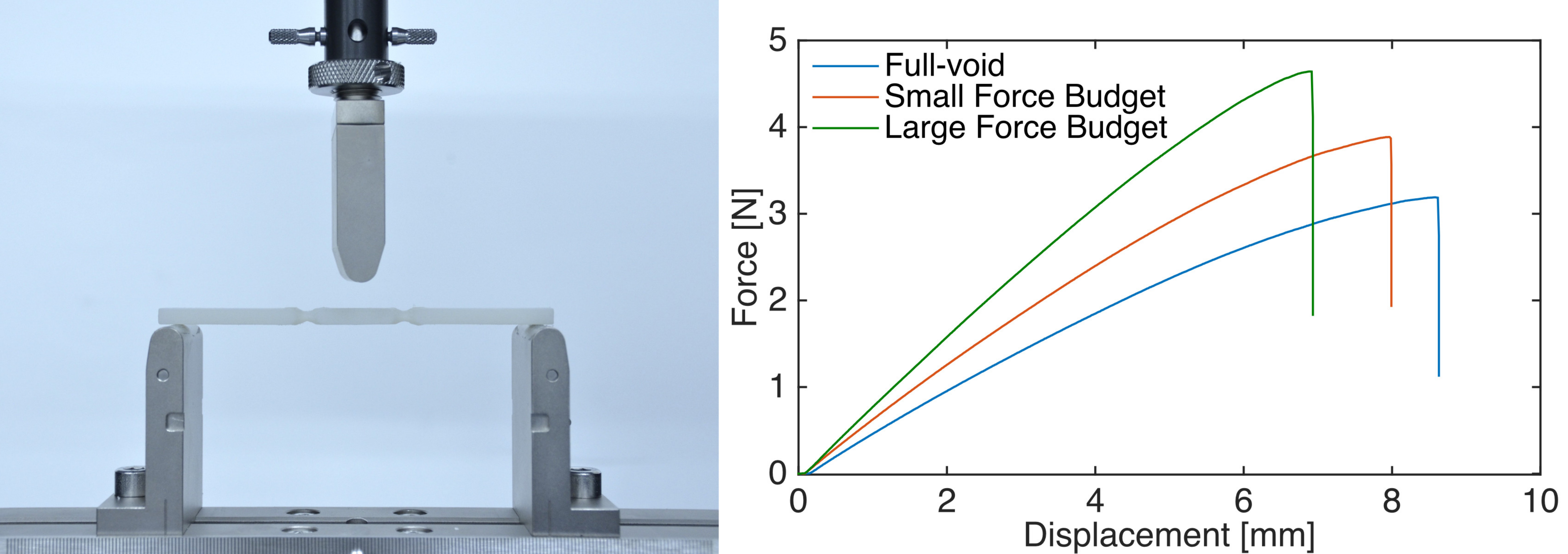}
\caption{Three point bend tests on optimized test models. The model optimized for larger force budget performs best while empty model breaks at a small force value. }
\label{fig:TestExperiment}
\end{figure}

Figure~\ref{fig:TestExample} illustrates a test model we designed to observe the effect of the force budget.
The model has two thin regions with slightly different dimensions.
We set the contact region to be the entire top surface while fixing it only at the bottom right and left edges to simulate a simply supported beam.
We then optimized the structure for two different force budgets of $4N$ and $5.5N$.
For the smaller force budget, the optimization focuses material around the thinnest part only.
However, for the larger force budget, material is distributed around both of the failure prone regions.
Notice that only a portion of the thinnest neck is filled for the small force budget while it is entirely filled for the larger force budget.
To validate the optimization results of the test model, we performed a set of three-point bend tests.
For benchmarking, we also tested a completely empty shell model (full-void). Figure~\ref{fig:TestExperiment} shows the experimental setup together with the force plots we obtained. The test model optimized for the larger force budget performs best while the empty model breaks at the smallest force magnitude.
Our structural optimization method strengthens the model up to $46\%$ while increasing its mass by only $15\%$.
The  discrepancies  between the input force budget  and the measured failure forces can be due to the FEA modeling of the problem as well as the possible anisotropic behavior of the 3D printed material \cite{ulu2015enhancing}.

\paragraph{Performance} Table~\ref{tab:Results} shows the performance of our algorithm.
We tested our method on a 3.2GHz Intel Core i5 computer with 8GB of memory.
We selected various 3D models and optimized under different force configurations.
Although the major computational cost comes from the critical instant analysis, we achieve $5\times$ acceleration on average  over a brute force approach. This acceleration becomes more significant as the contact region (hence the number of force instants) grows in relation to the total boundary surface. Shark (large contact region) vs. Bracket (small contact region) in Table~\ref{tab:Results} highlights this difference. The main reason is that the stiffness matrix is assembled and factorized only once at each optimization step and $\sigma_{cr}$ is computed by only performing a back-substitution for each force instant. For a large number of force instants, the cost of a single assembly and factorization becomes much smaller compared to the number of linear solves performed. 

\subsection{Limitations and Future Work}

Our critical instant analysis is based on an approximation to the relationship between input forces and resulting stresses. We found this approximation to work well when stresses due to bending and torsion are dominant compared to local compressive stresses. For shapes where local compressive stresses play a dominant role in failure,  an efficient approximation of stress and an accurate estimation of the criticality map  remains an open problem. For models with many small protruding features, our geodesic force instant sampling may fail to sample such features, thus causing the estimated criticality map to miss potential critical instants. 

In our approach, we benefit from the boundary shell in preserving the external shape and solving the singularity problems in the optimization. However, it can also serve as a main structural component as its thickness is increased. In this paper, we do not optimize the boundary shell thickness. A natural extension of our approach would be to efficiently determine the boundary shell thickness as a preprocessing step to our algorithm.

In the future, our analysis could be extended nonlinear and/or anisotropic material models. One of the advantages of using material modes is that the resulting density is smoothly varying, which makes the results easier to fabricate. This work focuses on making objects as safe at all times, but in some cases one may want to incorporate \textit{weak} points to enable fail-safe designs. Our critical instant analysis might be able to handle this case by using a criticality map construction that takes into account a spatially-varying thresholds.

\section{Conclusion}

We present a lightweight structure optimization method for 3D objects under force location uncertainty.
We propose a novel critical instant analysis method to efficiently determine the force instant creating the highest stress in the structure.
With this method, we show that an approximation to the relationship between the force configurations and  resulting stress distributions can be captured using only small number of FEA evaluations.
Combined with a reduced order formulation, we demonstrate that our method provides a practical solution to this computationally demanding  optimization problem.
We evaluate the performance of our algorithm on a variety of 3D models.
Our results show that significant mass reduction can be achieved by optimizing the material distribution while ensuring that the object is structurally sound against a wide range of force configurations capped by a force budget.

\appendix
\section{Analytic Gradients}
\label{sec:Appendix}

\myColor{Following the final formulation in \eqref{Eq:reducedOptimizationProblem}, the gradient of mass with respect to the reduced order design variables $\boldsymbol{\alpha}$ can be calculated as}

\begin{equation}
\myColor{ \frac{\partial{M}}{\partial{\boldsymbol{\alpha}}} = \frac{\partial{M}}{\partial{\boldsymbol{\rho}}} \frac{\partial{\boldsymbol{\rho}}}{\partial{\boldsymbol{\alpha}}}}
\end{equation}

\myColor{where the first term is simply the elemental volume vector $\boldsymbol{V}$ and the second term can be obtained by following \eqref{Eq:logistic} as}

\begin{equation}
\myColor{ \frac{\partial{\boldsymbol{\rho}}}{\partial{\boldsymbol{\alpha}}}  = \frac{\partial{G}}{\partial{\boldsymbol{x}}} \frac{\partial{\boldsymbol{x}}}{\partial{\boldsymbol{\alpha}}}.}
\end{equation}

\myColor{Here, $\boldsymbol{x} = \boldsymbol{\Gamma} \boldsymbol{\alpha}$ and its derivative with respect to $\boldsymbol{\alpha}$ is simply the reduced order material basis matrix $\boldsymbol{\Gamma}$. }

\myColor{The gradient of the critical stress $\sigma_{cr}$ with respect to $\boldsymbol{\alpha}$ can be obtained following the formulation in \eqref{Eq:reducedSigmaCritical} as}

\begin{equation}
\label{dsigmacr_dalpha}
\myColor{ \frac{\partial{\sigma_{cr}}}{\partial{\boldsymbol{\alpha}}} = \frac{\partial{\sigma_{cr}}}{\partial{\boldsymbol{\rho}}} \frac{\partial{\boldsymbol{\rho}}}{\partial{\boldsymbol{\alpha}}}.}
\end{equation}

\myColor{We use p-norm approximations ($p=15$) for the max functions. For $H(\boldsymbol{\sigma}^{vm}) = \| \boldsymbol{\sigma}^{vm} \|_{p}$ where $\boldsymbol{\sigma}^{vm}$ is composed of $\sigma_e^{vm} \quad \forall e \in \mathcal{V}_{wr}$, the derivative of $\sigma_{cr}$ with respect to $\boldsymbol{\rho}$ can be computed as }

\begin{equation}
\label{dsigmacr_dro}
\myColor{ \frac{\partial{\sigma_{cr}}}{\partial{\boldsymbol{\rho}}} = \frac{\partial{H}}{\partial{\boldsymbol{\sigma}^{vm}}} \frac{\partial{\boldsymbol{\sigma}^{vm}}}{\partial{\boldsymbol{\sigma}}} \left( \frac{\partial{\boldsymbol{\sigma}}}{\partial{\boldsymbol{u}}} \frac{\partial{\boldsymbol{u}}}{\partial{\boldsymbol{\rho}}} +  \frac{\partial{\boldsymbol{\sigma}}}{\partial{\boldsymbol{\rho}}} \right) }
\end{equation}

\myColor{where the derivative of $\boldsymbol{u}$ with respect to $\boldsymbol{\rho}$ can be obtained from the equilibrium equation $\boldsymbol{K}\boldsymbol{u} = \boldsymbol{f}$ as}

\begin{equation}
\label{du_dro}
\myColor{ \frac{\partial{\boldsymbol{u}}}{\partial{\boldsymbol{\rho}}} = \boldsymbol{K}^{-1} \left( -\frac{\partial{\boldsymbol{K}}}{\partial{\boldsymbol{\rho}}} \boldsymbol{u} \right). }
\end{equation}

\myColor{Applying the adjoint variable method, \eqref{dsigmacr_dro} can be re-written as}

\begin{equation}
\label{dsigmacr_dro_adjoint}
\myColor{ \frac{\partial{\sigma_{cr}}}{\partial{\boldsymbol{\rho}}} = \boldsymbol{\xi}^T \left(- \frac{\partial{\boldsymbol{K}}}{\partial{\boldsymbol{\rho}}} \boldsymbol{u} \right)  + \frac{\partial{H}}{\partial{\boldsymbol{\sigma}^{vm}}} \frac{\partial{\boldsymbol{\sigma}^{vm}}}{\partial{\boldsymbol{\sigma}}}\frac{\partial{\boldsymbol{\sigma}}}{\partial{\boldsymbol{\rho}}} }
\end{equation}

\myColor{where $\boldsymbol{\xi}$ is the adjoint variable and defined as }

\begin{equation}
\label{adjointVariable}
\myColor{ \boldsymbol{\xi}^T = \frac{\partial{H}}{\partial{\boldsymbol{\sigma}^{vm}}} \frac{\partial{\boldsymbol{\sigma}^{vm}}}{\partial{\boldsymbol{\sigma}}} \frac{\partial{\boldsymbol{\sigma}}}{\partial{\boldsymbol{u}}}\boldsymbol{K}^{-1}. }
\end{equation}

\myColor{All the terms in equations \eqref{dsigmacr_dalpha} and \eqref{dsigmacr_dro_adjoint} can be directly obtained and the adjoint variable can be computed by solving the system of linear equations in \eqref{adjointVariable}.}

\begin{acks}

\myColor{The authors would like to thank Stelian Coros and Paul Steif for insightful discussions; Kadri Bugra Ozutemiz, Robin Song and Nicholas Harrington for their help in physical testing and Recep Onler for his help in video production. We are grateful to the designers whose 3D models we used: Aim@Shape for the fertility, the Stanford Computer Graphics Laboratory for the Stanford bunny, thundermooo at Thingiverse for the chair, Sorkine-Hornung~\textit{et al.} for the cactus, Langlois~\textit{et al.} for the penguin, Lu~\textit{et al.} for the shark and Qingnan Zhou for the rocking chair. This work is partly funded by NCDMM America Makes Project \#4058.}


\end{acks}

\bibliographystyle{ACM-Reference-Format}
\bibliography{References}

\end{document}